# Hidden Magnetic Order in Triangular-Lattice Magnet $Li_2MnTeO_6$

E.A. Zvereva[1,7,*], G.V. Raganyan[1], T.M. Vasilchikova[1], V.B. Nalbandyan[2], D.A. Gafurov[3], E.L. Vavilova[3], K.V. Zakharov[1], H.-J.Koo[4], V.Yu. Pomjakushin[5], A.E. Susloparova[6], A.I. Kurbakov[6], A.N. Vasiliev[1,7,8], and M.-H. Whangbo[9,10]

[1]*Faculty of Physics, Lomonosov Moscow State University, Moscow, 119991, Russia*

[2]*Chemistry Faculty, Southern Federal University, Rostov-on-Don, 344090 Russia*

[3]*Zavoisky Physical-Technical Institute, FRC Kazan Scientific Center of RAS, Kazan, 420029 Russia*

[4]*Department of Chemistry and Research Institute for Basic Sciences, Kyung Hee University, Seoul 02447, Korea*

[5]*Laboratory for Neutron Scattering and Imaging LNS, Paul Scherrer Institute, Villigen, CH-5232 Switzerland*

[6]*NRC «Kurchatov Institute» - PNPI, Gatchina, 188300, Russia*

[7]*National Research South Ural State University, Chelyabinsk 454080, Russia*

[8]*National University of Science and Technology "MISiS", Moscow 119049, Russia*

[9]*Department of Chemistry, North Carolina State University, Raleigh, NC 27695-8204, USA*

[10]*State Key Laboratory of Structural Chemistry, Fujian Institute of Research on the Structure of Matter (FJIRSM), Chinese Academy of Sciences (CAS), Fuzhou 350002, China*

The manganese tellurate $Li_2MnTeO_6$ consists of trigonal spin lattices made up of $Mn^{4+}$ ($d^3$, $S$ = 3/2) ions. The magnetic properties of this compound were characterized by several experimental techniques, which include magnetic susceptibility, specific heat, dielectric permittivity, electron spin resonance (ESR), nuclear magnetic resonance (NMR) and neutron powder diffraction (NPD) measurements, and by density functional calculations (DFT). The magnetic susceptibility $\chi(T)$ demonstrates very unusual behavior. It isdescribed by the Curie-Weiss law at high temperature with Curie-Weiss temperature of $\Theta$ = -74 K, exhibits no obvious anomaly indicative of a long-range magnetic ordering at low magnetic fields. At high magnetic fields, however, the character of $\chi(T)$ changes showing a maximum at about 9 K. That this maximum of $\chi(T)$ reflects the onset of an antiferromagnetic order was confirmed by specific heat measurements, which exhibit a clear $\lambda$-type anomaly at $T_N \approx 8.5$ K even at zero magneticfield, and by $^7$Li NMR and dielectric permittivity measurements. The magnetic structure of $Li_2MnTeO_6$, determined by neutron powder diffraction measurements at 1.6 K, is described by the 120° non-collinear spin structure with the propagation vector $\boldsymbol{k}$ = (1/3, 1/3, 0). Consistent with this finding, the spin exchange interactions evaluated for $Li_2MnTeO_6$ by density functional calculations are dominated by the nearest-neighbor antiferromagnetic exchange within each triangular spin lattice. This spin lattice is strongly spin frustrated with $f = |\Theta|/T_N \approx 8$ and exhibits a two-dimensional magnetic character in a broad temperature range above $T_N$.

## I. INTRODUCTION

Systems staying in disordered spin-liquid-like state down to lowest temperatures are of great interest as candidates in which to discovernovel quantum phenomena [1-4]. The most promising candidates are low-dimensional frustrated magnets, which are prevented from achieving along-range order (LRO) by Mermin-Wagner theorem [5]. Though rare, there occurs an exotic situation in whicha system undergoes a long-range ordering but its observation is hampered [6-8]. In this case, the standard macroscopic magnetic techniques such as magnetic susceptibility, specific heat or even neutron scatteringmeasurements are insufficient for gaining insight into the ground state of a magnetically ordered solid, particularly if it exhibits a short-range orderor nearly-random order. Systems possessing this type of hidden magnetic order have not been wellstudied, and studieson such systems can lead to interesting results.

Historically, the heavy Fermion material $URu_2Si_2$ was the first compound for which a hidden (i.e., not observable by standard techniques) magnetic order was discovered [9, 10]. Numerous experimental efforts have been made to uncover the true nature of this phenomenon with tiny magnetic moments below a transition temperature. One of the recent explanations includes the formation of a nontrivial magnetic dotriacontapole (rank five) moments as the order parameter [7]. The cause forthe violation of the Stoner criteria in actinides is thought to arise from a strong spin-orbit coupling. In contrast to conventional magnets, therefore, the Fermi surface nesting stabilizes staggered multipole moments, which can be viewed as an unconventional density wave (UDW) [6, 7, 11].

In ferropnictide superconductors, the magnetic moments on iron cancel out below the quantum critical point. The hidden magnetic order behind this observation was related to a weak Hund's rule coupling, in contrast to a commensurate spin density wave (cSDW), which occurs at a strong Hund's rule coupling [12]. It was assumed that injected mobile holes can account for the nature of low-energy single-particle excitations in the hidden magnetic order states in ferropnictide materials.

There are a few remarkable examples of hidden magnetic order in low-dimensional magnets. In particular, an intricate magnetic transition around 70 K has been observed for quasi-one-dimensional (1D) spin lattice semiconducting magnet CuNCN [8]. Despite the lack of specific heat anomaly as well as the missing magnetic scattering in the neutron diffraction data, the antiferromagnetic (AFM) LRO was evidenced by a bending in the temperature dependence of the magnetic susceptibility and clearly established by the spectroscopic techniques like muon spin spectroscopy (μSR), ESR and NMR [13]. It has been shown that the experimental observation of the LRO has been impeded bystrong quantum fluctuations that reduce the Néel temperature and the ordered moment.

The coexistence of two spin sublattices with mixed dimensionalities [e.g., 1D and two-dimensional (2D)] and energy scales was believed to be the reason for the hidden magnetic order in another Cu-based compound $CuP_2O_6$ [14]. The ordering weakly manifests itself in the magnetic susceptibility but has not been seen by neutron diffraction, while ESR data confirmthe LRO onset with low net moment indicating substantial fluctuations that persist even below $T_N$.

An unusual hidden magnetic ordering occurs in aninsulating oxide $Sr_2VO_4$ at low temperature [15-19], for whichseveral possible scenarios have been mentioned including an orbital ordering triggered by Jahn-Teller distortion [15], an unconventional magnetic octupolar ordering [20], and a Néel order with muted order parameter of 0.06 $\mu_B$(compared to 1 $\mu_B$ in a classical spin-1/2 systems) [16, 19]. A direct confirmation of the LRO and inhomogeneous magnetic state was obtained by μSR studies [17], while recent theoretical calculations predict the single-stripe magnetic ordering [18].

In the present article, we examine the properties of a quasi-2D triangular lattice compound $Li_2MnTeO_6$, which belongs to the new $A_2MnTeO_6$ (A = Li, Na, Ag, Tl) structural family of frustrated low-dimensional magnets [21], to find that the magnetic properties of $Li_2MnTeO_6$ are similar to those magnets possessing unusual hidden magnetic order.

## II. EXPERIMENTAL

Powder samples of $Li_2MnTeO_6$ were prepared by a molten-salt ion exchange reaction of the isostructural $Na_2MnTeO_6$, as reported previously [21]. $Li_2MnTeO_6$ crystallizes in the space group $P\bar{3}1c$. The $Mn^{4+}$ and $Te^{6+}$ cations of $Li_2MnTeO_6$ form magnetic honeycomb layers, which alternate with nonmagnetic layers of $Li^+$ cations. Each individual $(MnTeO_6)^{2-}$ layer is essentially ordered such that the magnetic $Mn^{4+}$ions form a triangular networkconsisting of weak spin exchange paths Mn-O-Te-O-Mn (Fig. 1). Thus, the crystal structure providesthe conditions necessary for low-dimensional frustrated magnetic interactions.

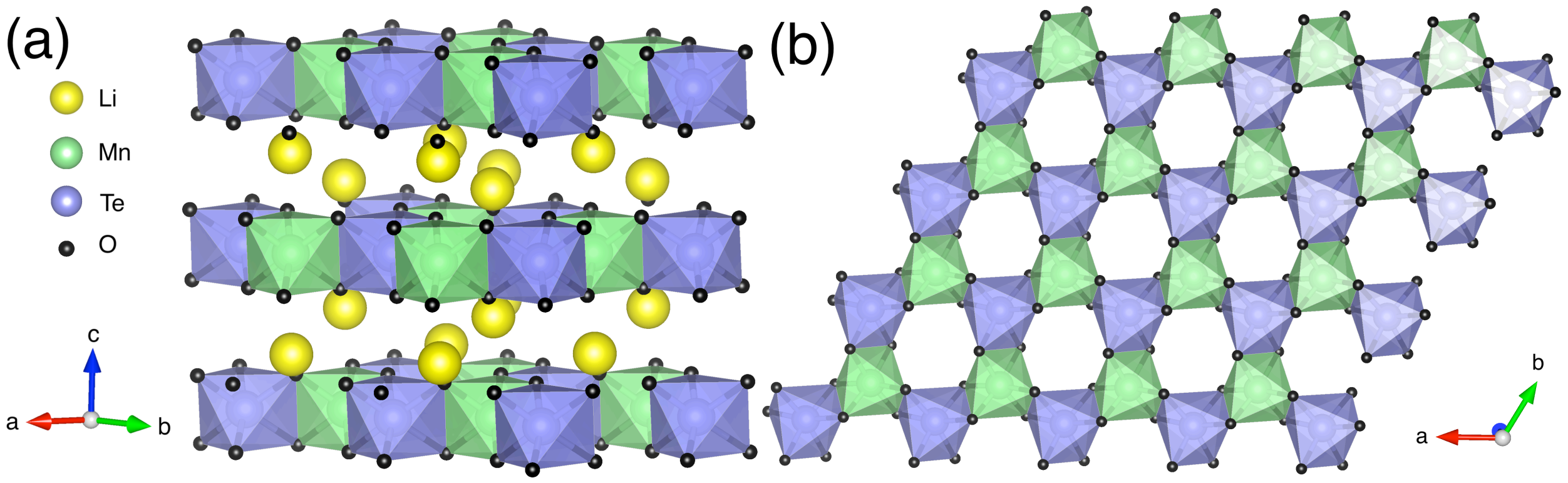

FIG. 1.Polyhedral view of the layered crystal structure of $Li_2MnTeO_6$ ($P\bar{3}1c$ space group): (a) A perspective view of layers. (b) A projection of the layer of magnetic ions along the *c*-direction, in which the $MnO_6$ and $TeO_6$ octahedra alternate sharing their edges. The green and blue octahedra represent the $MnO_6$ and $TeO_6$, respectively. The Li and O atoms are represented by yellow spheres and black balls, respectively.

The temperature-dependent magnetic susceptibilities were measured at the magnetic field $B$ = 0.1, 1, 3, 6 and 9 T in the temperature range 2−300 Kby means of a Quantum Design PPMS systemusing VSM Teflon capsule with 34 mg of powder sample. Specific heat measurements were carried out by a relaxation method using a Quantum Design PPMS system on a cold-pressed 5.03 mgsample. The data were collected at zero magnetic field as well as under applied fields of 3, 6 and 9 T in the temperature range 2−300 K. The dielectric constant was measured on a Quantum Design MPMS XL-7 system using a custom-made insert and Andeen-Hagerling 2700A capacitive bridge. The measurements were performed at various frequencies in the range 1 – 20 kHz range at temperatures from 2 to 300 K in zero magnetic field. The resolution of the Andeen-Hagerling 2700A bridge is 2.4 - 16 aF depending on the frequency. In addition, the device can measure the loss tangent up to $1.5 \times 10^{-8}$, the conductivity down to $3 \times 10^{-16}$S, or the resistance up to $1.7 \times 10^{15}$ Ohm, with the operating voltage of 15 V. The duration of one measurement is from 30 ms to 0.4 s. The sample for dielectric permittivity measurements was a plane-parallel disk with a diameter of up to 6 mm and a thickness of up to 3 mm. A silver paste, "LietSilber", was applied to both sides of the disk, resulting in a capacitor. The sample was mounted in the measuring insert for the MPMS XL-7 system, which was used to control the temperature and external magnetic field.

Electron spin resonance (ESR) studies were carried out using an X-band ESR spectrometer CMS 8400 (ADANI) ($f \approx 9.4$ GHz, $B \leq 0.7$ T) equipped with a low-temperature mount, operating in the 6−270 K range. The effective g-factor of our sample was calculated with respect to an external reference for the resonance field. We used a,g-bisdiphenyline-*b*-phenylallyl (BDPA) ($g_{et}$ = 2.00359) as a reference material. Nuclear magnetic resonance (NMR) measurements were carried out for $^7$Li nuclei (spin $I = 3/2$, gyromagnetic ratio $\gamma_n$= 16.55 MHz/T) at constant frequencies 13.0 and 116.56 MHz in the temperature range 4.2-100 K. The data were collected using a Tecmag Redstone pulse solid-state NMR spectrometer. For recording the spectra, we used a solid-echo sequence as the main experimental technique. The magnetic field varied in the range of 0.59 − 0.99 T and 6.87 − 7.22 T with a step from 0.01 to 0.0005 T depending on the measurement temperature. The relaxation rates were measured by the saturation recovery method. The nuclear magnetization $M$ versus the pulse delay time $\tau$ was fitted to a single exponential function $M(\tau) = M0\{1 - f\exp[-(\tau/T_1)^b]\}$ with $0.7 < f < 0.9$ ($f = 1$ for ideal saturation). The constant $b = 1$ at $T > 10$ K but decreases slightly at low temperatures indicating the growth of electron spin system anisotropy. In principle, the main line of spin = 3/2 nuclei is described by a double-exponential dependence. In the present case, however, a single-exponential dependence describes the experiment more correctly, since the line shows a stronginhomogeneousbroadening and since quadrupole satellites are not resolved. Dynamic studies (ESR and NMR) were carried out using powder samples of 38 and 43 mg, respectively.

NPD measurements were carried out at the spallation neutron source SINQ at PSI Villigen using the two instruments, DMC (Cold Neutron Powder Diffractometer) and HRPT (High-Resolution Powder Diffractometer for Thermal Neutrons). The HRPT diffractometer uses thermal neutrons with $\lambda$ = 1.886 Å, monochromatized by a focusing single crystal germanium monochromator with Ge (335) reflection. The fixed take-off monochromator angle was 120°. Neutron diffraction patterns were collected in the 2θ diffraction angle range of $3.55^0 - 164.50^0$ with steps of $0.05^0$. The sample was placed into a thin-walled cylindrical vanadium container with dimensions 6 mm × 50 mm. HRPT measurements at room temperature were carried out to verify the quality of the sample and refine the crystal structure. The latter is possible due to the high resolution of the diffractometer. Measurements at the lowest achieved temperature, $T$ = 1.6 K, were carried out to record magnetic neutron scattering associated witha long-range magnetic order. This allows one to construct and describe the spin structure in the ground state.

To analyze the temperature evolution of the magnetic scattering, neutron diffraction patterns were measured on a high intensity DMC powder diffractometer using cold neutrons monochromatized with a graphite monochromator. The measurements were carried out at a wavelength, $\lambda$ = 2.4575 Å, of the neutron beam incident on the sample in the angular range of $5.0^0$ - $86.7^0$ with a step of $0.1^0$. The sample neutron diffraction patterns were recorded at temperatures $T$ = 1.6, 2.5, 4, 6, 10, 20 K. All diffraction patterns from HRPT and DMC were treated with the Rietveld method using the FULLPROF suite [22].

## III. RESULTS AND DISCUSSION

### A. Sample characterization

The neutron diffraction pattern measured on the HRPT at room temperature and the results of itsRietveld refinement are presented in Fig. 2. It clearly demonstrates the high quality of the sample under study and the absence of any meaningful impurities. All diffraction peaks, even low-intensity ones, are well described in terms of the space group $P\bar{3}1c$ ($a = b$ = 5.01710(6) Å, $c$ = 9.5267(2) Å at $T$ = 300 K). The narrow and symmetrical shapes of the diffraction maxima indicate a high crystallinity of the sample. A microstructural analysis using the Rietveld refinement has slightly improved the quality of the fitdue to the introduction of microstrains.

There is a significant difference in magnitude and even sign in the neutron scattering lengths of the elements constituting the compound (i.e. $b_{Li} = -0.19\times10^{-12}$ cm, $b_{Mn} = -0.37\times10^{-12}$ cm, $b_{Te} = 0.82\times10^{-12}$cm, and $b_O = 0.58\times10^{-12}$cm). Thus, neutron diffraction measurements are especially important for the analysis of the possible substitutional defects, which often happens in similar layered oxides.

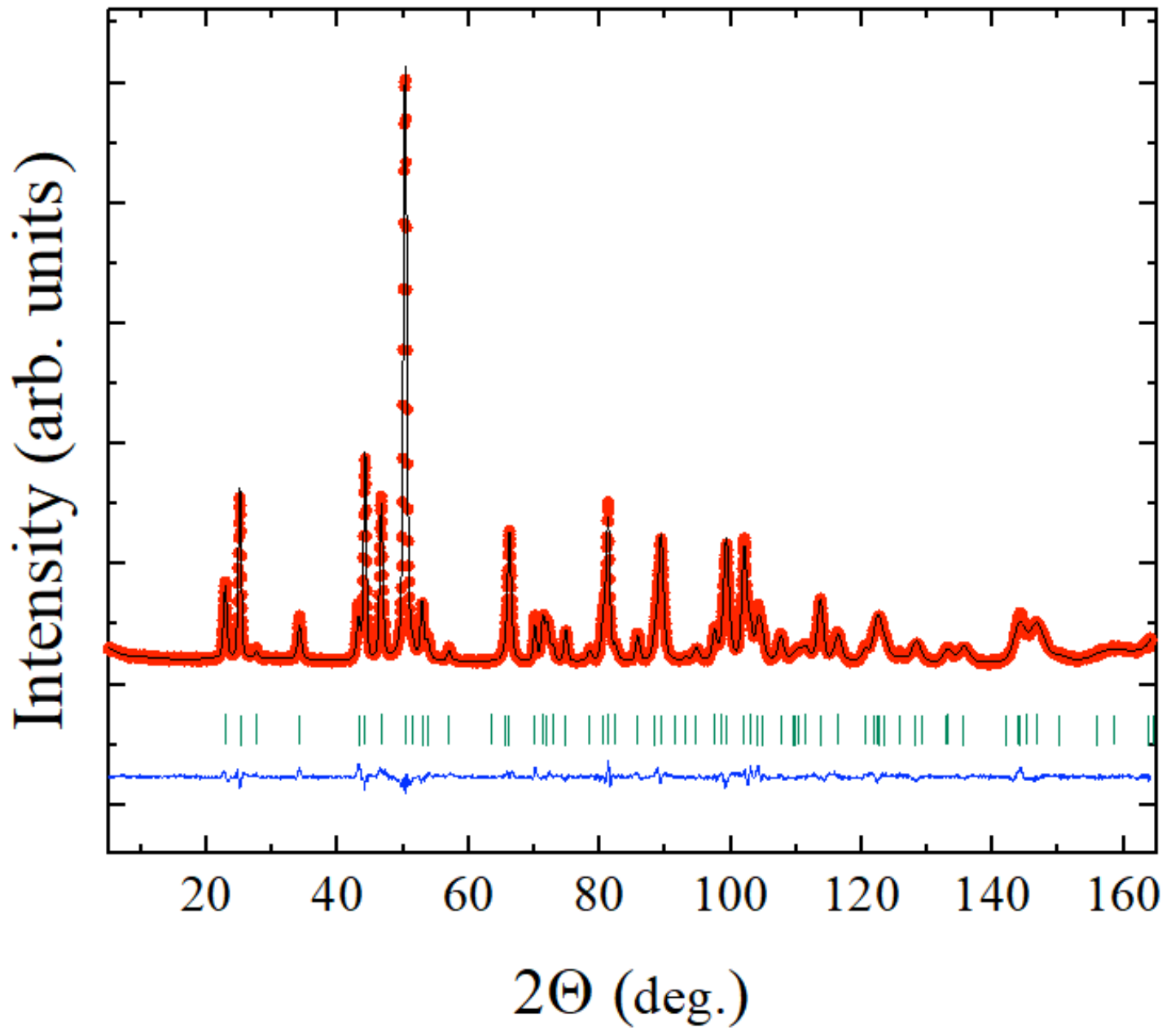

FIG. 2. Refined neutron diffraction pattern of $Li_2MnTeO_6$ measured on the HRPT diffractometer at $T$ = 300 K. The red dots represent the experimental data, the black line shows the calculated intensity, the green ticks indicate the position of the Bragg reflections, and the blue line shows the difference between experimental and calculated data, which is plotted at the bottom for convenience.

## B. Static magnetic properties and dielectric permittivity

The temperature dependence of the magnetic susceptibility $\chi = M/B$ of $Li_2MnTeO_6$, measured in the field $B$ = 0.1 T, is shown in Fig. 3. The zero-field-cooled (ZFC) and field-cooled (FC) susceptibilities do not show noticeable divergence indicating the absence of impurities and spin-glass effects. This is in good agreement with the neutron data, which indicate a high purity of the sample. In weak magnetic fields, the behavior is almost a Curie-Weiss type without any clear anomaly that can beassociated with a long-range magnetic order. It is found, however, that an increase in the probe magnetic field strength leads to significant changes in the behavior of the temperature dependence of the susceptibility with the appearance of a maximum of $M(T)$ at approximately $T_{max} \approx 9.2$ K (inset in Fig.3). In the paramagnetic state, the $\chi(T)$ can be approximated by a modified Curie–Weiss law with $\chi = \chi_0 + C/(T- \Theta)$, where $\chi_0$ is the temperature-independent term, $C$ is the Curie constant, and $\Theta$ is the Weiss temperature.

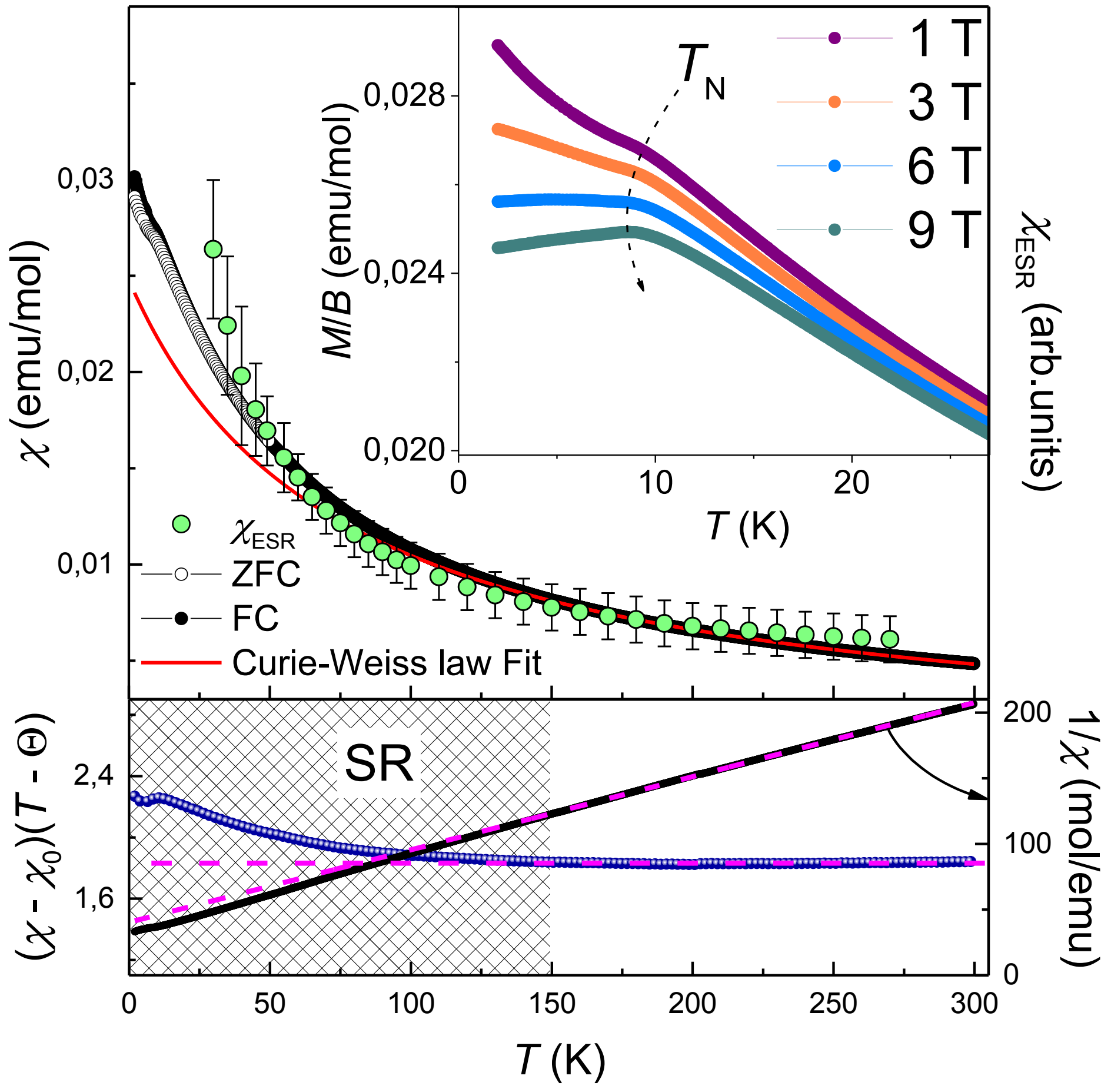

FIG. 3. The temperature dependence of the magnetic susceptibility $\chi = M/B$ for $Li_2MnTeO_6$ at $B$ = 0.1 T in both FC and ZFC modes (upper panel) and the integrated ESR intensity (green symbols). The inset to this panel represents the $M/B(T)$ curves for the $Li_2MnTeO_6$ at various external magnetic fields. The solid red curve represents the Curie-Weiss law. The temperature dependence of the Curie constant $C = (\chi - \chi_0)(T - \Theta)$ is shown in the lower panel along with the inverse magnetic susceptibility $1/\chi$. The dashed line represents the limiting value $C$ = 1.84 emu/(molK).

The diamagnetic contribution $\chi_{dia}$ was determined by summing up the Pascal constants [23] and was fixed to reduce the number of variable parameters in the fitting analysis, which yields $C$ = 1.84 emu K/mol and $\Theta$ = -74 ± 1 K. The obtained Curie constant $C$ gives the value of the effective magnetic moment as $\mu_{eff} = \sqrt{8C}\,\mu_B$ = 3.84$\mu_B$, which is in good agreement with theoretical estimation $\mu_{theor} = \sqrt{g^2S(S+1)}\,\mu_B$= 3.87$\mu_B$ for $Mn^{4+}$ ($S$ = 3/2). The Curie-Weiss temperature $\Theta$ is negative and relatively large in comparison with the $T_{max}$ (≈ 9.2 K), indicating the presence of dominant antiferromagnetic exchange interactions. With the frustration index $f = |\Theta|/T_N \approx 8$, the extent of spin frustration in the spin lattice isappreciable. In addition, at temperatures below ~ 150 K, a noticeable deviation of the $\chi(T)$ from the Curie-Weiss law is observed.This indicates the occurrence of strong short-range exchange interactions and a nontrivial ground state (lower panel of Fig. 3).

Despite the absence of a sharp anomaly in the magnetic susceptibility in weak magnetic fields, the low-temperature data on the specific heat unambiguously indicate the onset of a long-range magnetic order in $Li_2MnTeO_6$. The data on the specific heat $C_p(T)$ in a zero magnetic field exhibit a distinct $\lambda$-type anomaly, which obviously corresponds toa long-range ordering transition (Fig. 4). The Néel temperature found from the specific heat data is $T_N$ = 8.5 K.

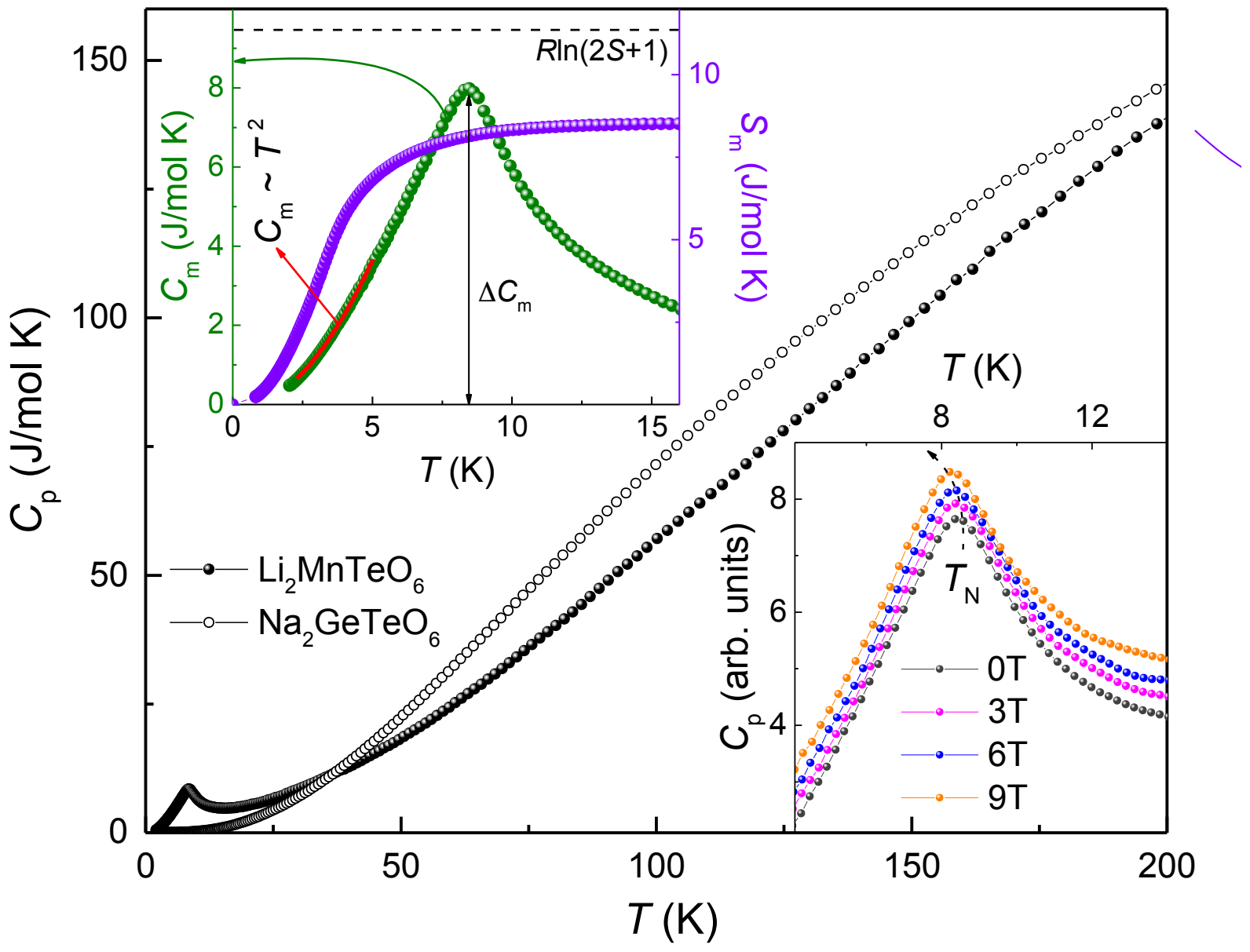

FIG. 4. The specific heat of $Li_2MnTeO_6$ (filled symbols) and that of non-magnetic isostructural analogue $Na_2GeTeO_6$ (open circles) under zero magnetic field. The upper inset shows the magnetic specific heat $C_m(T)$ (green symbols) and the magnetic entropy $S_m(T)$ (violet symbols). The red line is the result of the spin wave theory. The lower inset shows $C_p(T)$ under various magnetic fields, where the data were shifted upwards for clarity.

This value is slightly smaller than $T_{max}$ determined from the $M(T)$, which is typical of antiferromagnets with short-range interactions [24, 25].Under magnetic fields, the position of the $\lambda$-type anomaly in$C_p(T)$ shifts very slowly towards lower temperatures with increasing the field strength (the lower right inset of Fig. 4).

In order to clarify the nature of the magnetic phase transition and evaluate its effect on the specific heat and entropy, the temperature dependence of specific heat $C_p(T)$ was also measured for the isostructural nonmagnetic compound $Na_2GeTeO_6$. The Debye temperatures are estimated as $\Theta_D^{nonmag} = 421 \pm 5$K for $Na_2GeTeO_6$ and $\Theta_D^{mag} = 470 \pm 5$ K for $Li_2MnTeO_6$, after taking into account the molar mass differences between Li and Na and between Mn and Ge [26]. The standard scaling procedure [27] was applied to the $C_p(T)$ data for $Li_2MnTeO_6$ and $Na_2GeTeO_6$ to calculate the lattice phonon contribution $C_{ph}(T)$. The resulting $C_{ph}(T)$ was used to estimate the magnetic contribution $C_m(T)$ in $Li_2MnTeO_6$. We observe the specific heat jump $\Delta C_m = 8.3$ J/(mol K), which is lower than the value expected from the mean field (MF) theory because for the high-spin state of manganese $Mn^{4+}$ ($S = 3/2$), $\Delta C_{theor} = 5RS(S+1)/[S^2+(S+1)^2] = 18.3$ J/(mol K). The magnetic entropy $\Delta S_m$, collected up to $T_N$ is less than 50% of the total expected from the mean-field theory for a $S = 3/2$ spin system, i.e.,$\Delta S_m = R\ln(2S + 1) = 11.52$ J/(mol K) [26].

We also analyzed the magnetic contribution to the specific heat $C_m(T)$ below $T_N$ in terms of the spin-wave (SW) approach assuming that, at low temperature,the magnetic specific heat should follow the$C_m \approx T^{d/n}$ power law for magnons [28, 29], where the constant $d$ stands for the dimensionality of the magnetic lattice, and $n$ the exponent of the dispersion relation $\omega \approx \kappa^n$. For $Li_2MnTeO_6$, the least-squares fitting of the data below $T_N$ (the red solid line in theupper inset of Fig. 4) leads to the values of $d = 2 \pm 0.1$ and $n = 0.99 \pm 0.1$, which indicatesthe presence of 2D AFM magnons at very low temperatures.

The magnetic phase diagram can be constructed from our magnetic and thermodynamic data as depicted in Fig. 5. The long-range ordering into an AFM phase occurs at 8.5 K. This phase boundary is shifted slightly towards lower temperatures by magnetic field. As shown by the neutron diffraction studies and DFT calculations (see below), the magnetic structure is a 120°non-collinear spin arrangement expected for a triangular spin lattice with nearest-neighbor AFM spin exchanges (see below).

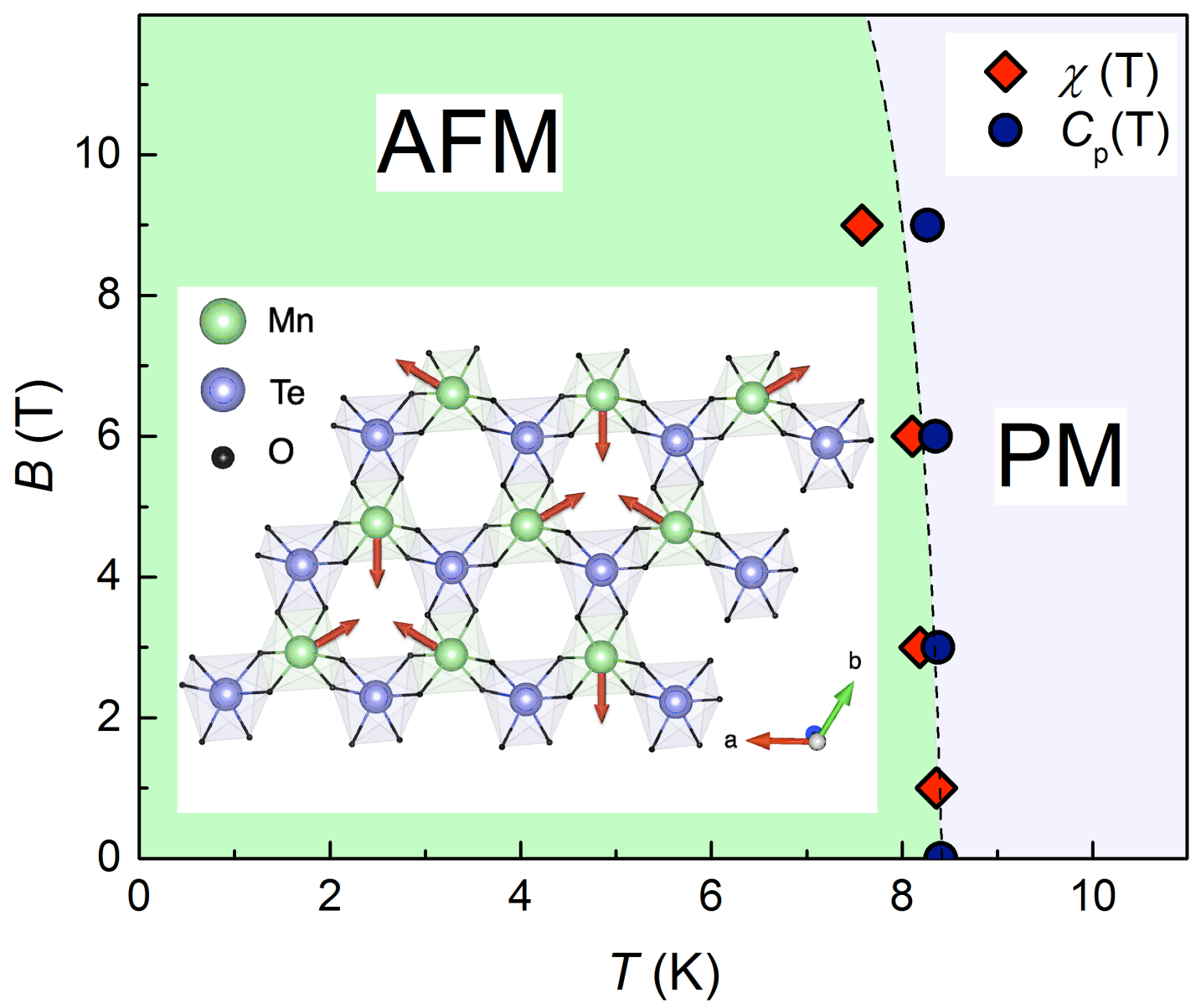

FIG. 5. The magnetic phase diagram for $Li_2MnTeO_6$. The inset shows the arrangement of magnetic sublattice.

The results of our dielectric measurements for $Li_2MnTeO_6$ reveals the presence of an additional anomaly in the vicinity of the ordering temperature as can be seen from Fig. 6, where the real part of dielectric permittivity is compared with specific heat and NMR data. Although this effect is weak, its presence is reproduced for all frequencies studied from 1 to 10 kHz. This fact may reflect the coupling of dielectric and magnetic subsystems in $Li_2MnTeO_6$ similar to that observed in triangular lattice antiferromagnets with non-collinear magnetic structure [30-33].

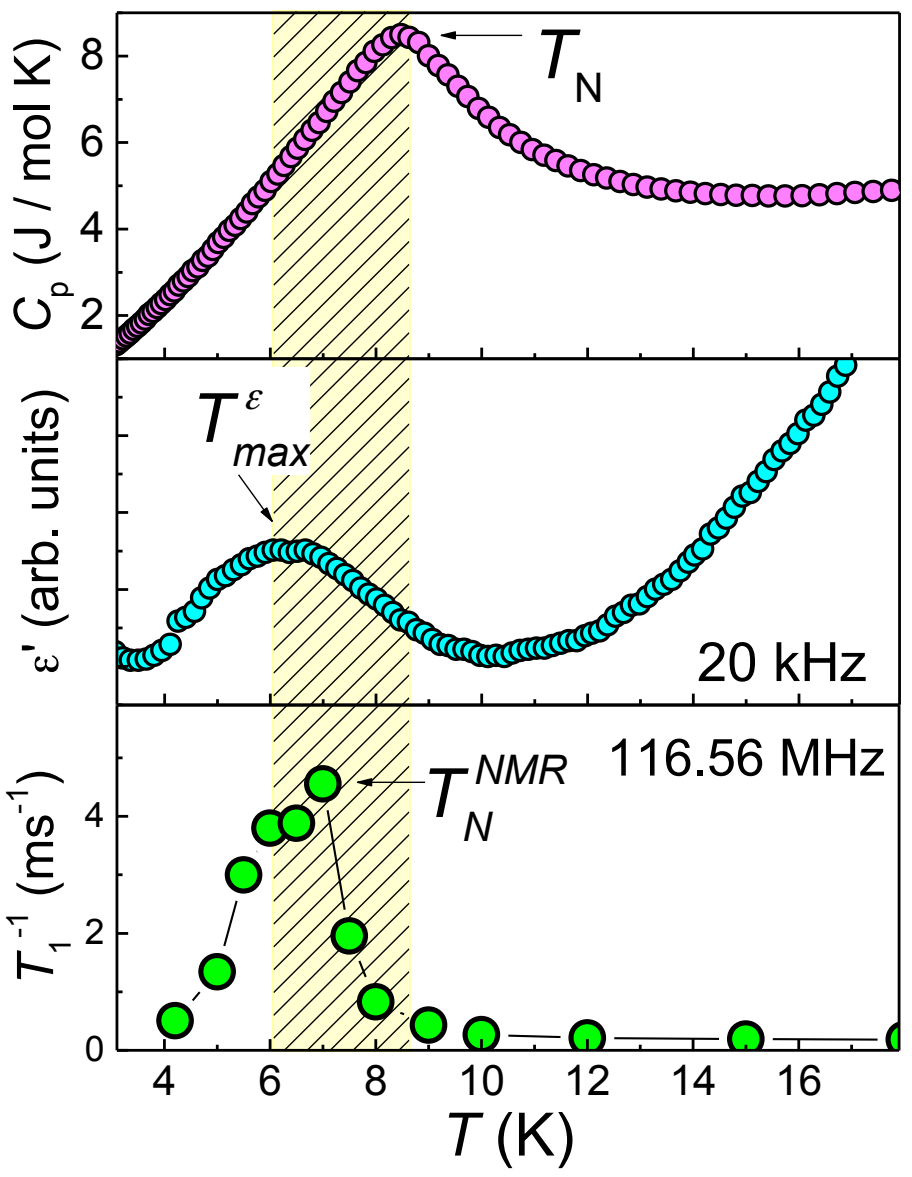

FIG. 6. Temperature dependence of the specific heat at zero magnetic field (upper panel), the real part $\varepsilon'$ of the permittivity recorded at frequency $f = 20$ kHz (middle panel) and the relaxation rate obtained at 116.56 MHz (lower panel), where the yellow area highlights the ordering transition region.

### C. Dynamic magnetic properties

Evolution of the ESR spectra with a temperature variation for powder sample of $Li_2MnTeO_6$ is shown in Fig. 7(a). In the entire range of temperature investigated, a relatively wide Lorentz-type line is observed, apparently corresponding to the signal from $Mn^{4+}$ ions. The integrated ESR intensity $\chi_{ESR}$, obtained by the double integration of the first derivative absorption line, was shown in the upper panel of Figure 2 along with static magnetic susceptibility. The shape of the line can be accounted for by fitting with a Lorentzian profile. Since the absorption line is relatively wide, we used the fitting expression that includes two circulating components of a linearly polarized high-frequency field [34]:

$$\frac{dP}{dB} \propto \frac{d}{dB}\left[\frac{\Delta B}{\Delta B^2 + \left(B - B_r\right)^2} + \frac{\Delta B}{\Delta B^2 + \left(B + B_r\right)^2}\right] \tag{1}$$

where $P$ is the power absorbed in the ESR experiment, $B$ the magnetic field, $B_r$ the resonant field, and $\Delta B$ the line width. The representative result of the fitting is shown by the red solid curve in the inset on Fig. 7(a). The effective g-factor and the ESR linewidth $\Delta B$ obtained from this approximation are collected in Fig. 7(b). It is clearly seen that the line is noticeably broadened with decreasing temperature, and the effective g factor deviates from a constant value below ~ 150 K, indicating an extended range of the short-range order correlations, which is typical for low-dimensional and frustrated systems [29]. In order to characterize the divergence of the temperature dependence of the linewidth, we have analyzed it in terms of the Dormann and Jaccarino [35] model and the classical critical behavior that, as the critical temperature is approached from above, the $\Delta B$ increases due to the slowing down of spin fluctuations [36-38]. In the paramagnetic regime far above the critical regions associated with the magnetic transition, where any critical contribution to $\Delta B$(T) as well as spin-lattice relaxation can be neglected, the temperature variation of the ESR linewidth can be described by the expression of Dormann and Jaccarino [35]:

$$\Delta B(T) = \Delta B_0 \cdot \left[\frac{\chi_0(T)}{\chi(T)}\right] \tag{2}$$

where $\chi_0 = C/T$ is the free single-ion susceptibility with $C$ the Curie constant of the uncoupled paramagnetic system, while $\chi(T)$ is the static susceptibility of the coupled system and $\Delta B_0$ describes the temperature-independent high-temperature limit of the linewidth associated with the contribution of anisotropic spin-spin interactions. We have calculated the temperaturevariation of $\Delta B$ using the measured dc susceptibility $\chi(T)$, the Curie constant $C$= $N_A g^2 \mu_B 2S(S+1)/3k_B$ = 1.84 emu/K mol of $Li_2MnTeO_6$ (with $S$ = 3/2 and experimental $g$ = 1.97 for $Mn^{4+}$ ions) as well as the constant value $\Delta B_0 \approx 90$ mT estimated in high temperature limit. As can be seen, the static susceptibility may roughly account for the temperature evolution of the $\Delta B$ (magenta dashed curve in Fig. 7b). As temperature decreases, however, appreciable deviation occurs from the experimental points.

In the Kawasaki-Mori-Huber theory [36-38], the temperature variation of $\Delta B$ can be described as $\Delta B(\mathrm{T}) = \Delta B^* + A(T^{\mathrm{ESR}}_{\mathrm{N}}/(T - T^{\mathrm{ESR}}_{\mathrm{N}}))^{\beta}$. Here A is an empirical parameter, $\Delta B^*$ is the high-temperature limiting value of the linewidth, $T^{\mathrm{ESR}}_{\mathrm{N}}$ the temperature of the order-disorder transition, and $\beta$ a critical exponent. A least squares fitting of the experimental data (the orange line in Fig.7(b)) reveals that the value of $T^{\mathrm{ESR}}_{\mathrm{N}} = 11 \pm 2$ K is close to the Néel temperature $T_{\mathrm{N}}$ obtained from the measurement of the specific heat. The critical exponent $\beta = 0.90 \pm 0.05$ K indicates the 2D nature [36] of the exchange correlations in $Li_2MnTeO_6$.

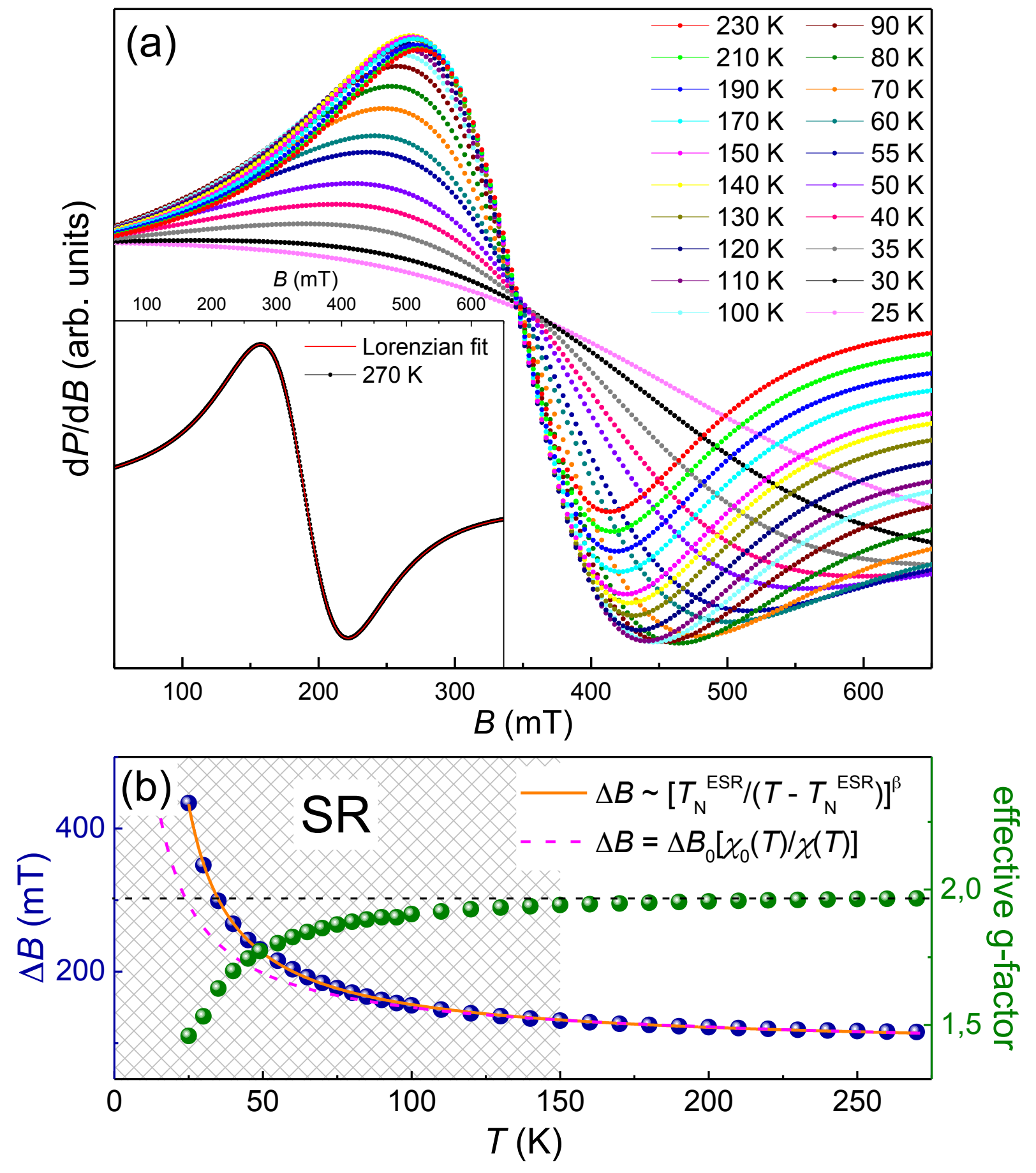

FIG. 7. (a) The temperature-dependent ESR spectra of $Li_2MnTeO_6$. The inset shows a representative ESR spectrum, wherethe Lorentzian fitting profile is given by thered solid line. (b) The temperature dependence of the effective *g* factor (green symbols) and the width of the ESR line (navy symbols). The solid orange curve corresponds to the temperature dependence of the linewidth within the framework of classical critical behavior and the magenta dashed line represents the prediction of Eq. 2.

At high temperatures, the $^7$Li NMR spectrum has a powder profile with unresolved quadrupole satellites. With decreasing temperature, the line shifts toward a lower field, and its width slightly increases. The local magnetic fields $H_{loc}$ cause a shift of the NMR line from the Larmor frequency $\omega_L = (\gamma_n/2\pi)H_0$, so the local electron spin susceptibility can be directly extracted from this shift. The shift is defined as $K=(\omega-\omega_L)/\omega_L \cdot 100\%$, and consists of twoterms, namely, $K(T)=K_{sp}(T)+K_0$, where $K_0$ is the temperature independent part of the shiftrelated to the Van Vleck susceptibility and the second-order quadrupolar effects, which are very weakly dependent on temperature (Fig. 8(a)). $K_{sp}$ is the spin part of the shift, caused by interactions $\hat{\mathcal{H}}_s = -\hat{I}\hat{A}\hat{S}$ between a nuclear spin $\hat{I}$ and an electron spin $\hat{S}$. It consists of three contributions $K_s = K_{dip} + K_{contact} + K_{core}$, where the Fermi contact term $K_{contact}$ and the core polarization term $K_{core}$ contribute to the isotropic part of the NMR shift while the dipolar term $K_{dip}$ depends on the relative orientation of the crystal axes to the external magnetic field and is therefore anisotropic. In the paramagnetic regime, $\hat{S}$ can be replaced by its expectation value $\langle S \rangle = \chi_s H_0$, and $K$is related to$\chi_{sp}(T)$ as

$$K_{sp}(T) = A_{hf}\chi_{sp}(T)/N_A\mu_B \quad (3)$$

where $A_{hf}$ is the hyperfine coupling constant, $N_A$ the Avogadro number, $\chi_{sp}$ the static spin susceptibility, and $\mu_B$ the Bohr magneton. Here we neglect the effect of the shape of the magnetic particles in the external field [39], which is much smaller than the transferred hyperfine field since the powder particles are almost spherical. The inhomogeneous linewidth is determined by the distribution of the local magnetic fields $H_{loc}$, so in paramagnetic regime its temperature change is also proportional to the static spin susceptibility $\chi_{sp}$.

The shift and width of the line obtained at $H_0 \approx 0.789$T depend linearly on the bulk static susceptibility $\chi$(T) obtained at 1 T in the temperature range 15 - 100 K (Fig. 8(b)). The hyperfine interaction constant $A_{hf}$ is determined to be 0.123 ± 0.009 kOe/$\mu_B$. It is quite small compared to the compounds with three-dimensional (3D) magnetic lattice (see for example $A_{hf}$ = 1.01 kOe/$\mu_B$ in spinel $LiMn_2O_4$ [40]). It is interesting that the obtained $A_{hf}$ value is similar to, or even smaller than, the parameter $A_{ax}$ = 0.24 ± 0.007 kOe/$\mu_B$, which is estimated using the point moment model of dipole-dipole interaction within the coordination radius of 6.5 Å. This suggests that, in layered compound $Li_2MnTeO_6$, the interplanar exchange interaction associated with the hybridization of orbitals and the transfer of spin density is very weak and partially compensates the local dipole field on the lithium nuclei due to the contact part of the hyperfine interaction. Below 15 K the line shift as well as the linewidth deviate strongly from the bulk static susceptibility. It means that the relation, $\tau_e^{-1} \gg \omega_L$, is not satisfied ($\tau_e$ is a correlation time of electron spins), and cannot be replaced any more by $\langle S \rangle$. This indicates the development of anisotropic electron correlations in the vicinity of the Néelphase transition temperature $T_N^{NMR} \approx$ 7K.TheNMR measurements are carried out in magnetic fields of several Teslas; therefore, the Néel temperature according to NMR is a bit lower than the one deduced from the zero field heat capacity data.

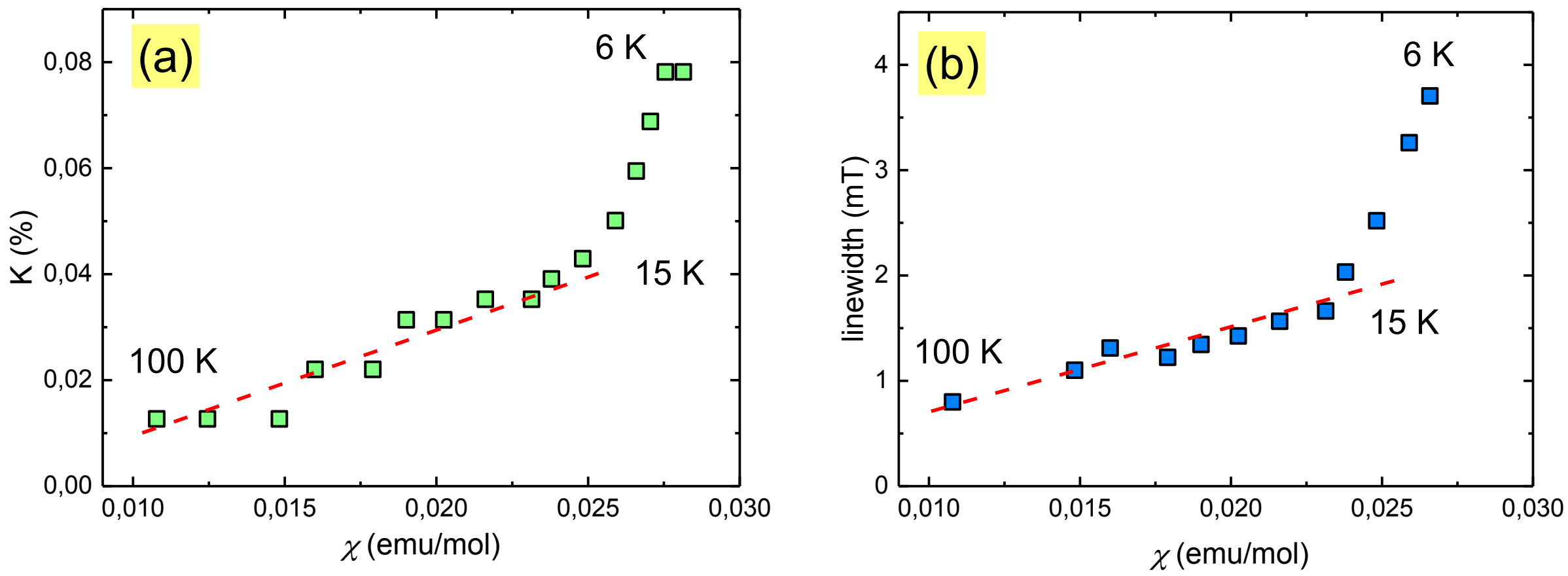

FIG. 8. (a) Line shift and (b) linewidth of the $^{7}$Li NMR signal at 13.05 MHz as a function of the static magnetic susceptibility. The dashed lines show a linear dependence of the local counterpart from the bulk static susceptibility.

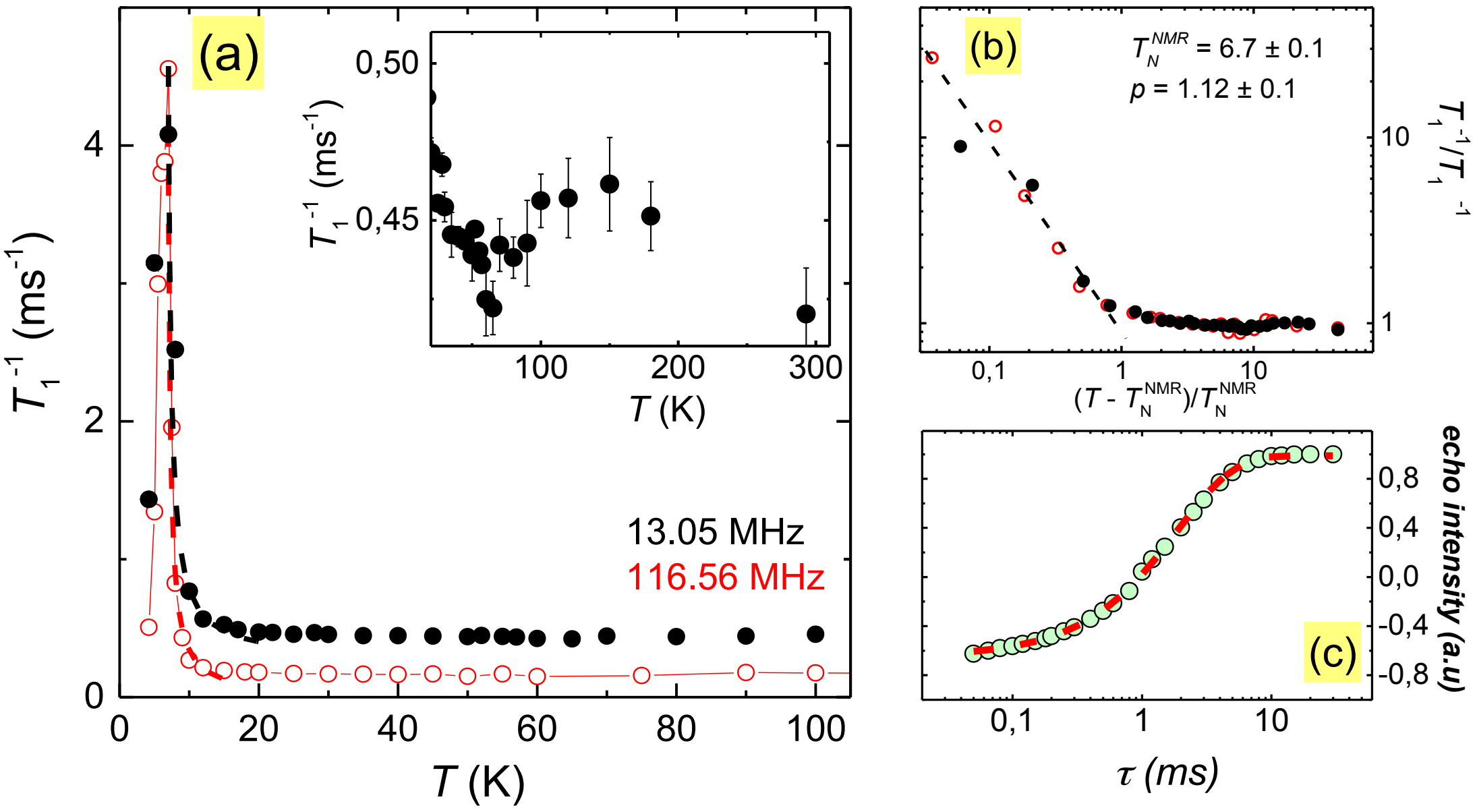

FIG. 9. (a) Temperature dependence of the relaxation rate obtained at 13.05 MHz (black solid circles) and 116.56 MHz (red open circles). The solid lines are guide for the eye. The inset shows high temperature region at 13.05MHz. (b) A log-log plot of the normalized $(1/^{7}T_1)/(1/^{7}T_{1\infty})$ against the reduced temperature $|T - T_N^{NMR}|/T_N^{NMR}$ for both frequencies ($T_N^{NMR}$ = 6.7 K). The dashed line represents eq. (5) with the critical exponent of $p$ = 1.12. (c) Example of an inversion recovery measurement on 13.05MHz at $T$ = 15K. Red line shows a fit by single exponent law (see text in part II).

The dynamics of the electron spin system was probed by the nuclear relaxation measurements. In the temperature range of 17 – 100 K the relaxation rate changes very slightly. Below 15 K, it grows sharply and has a peak at 6.7 K indicating the transition to an ordered state (Fig. 9). The temperature dependence of the spin-lattice relaxation observed in 7.045 T also shows a pronounced peak at 7 K.The relaxation process is dominated by the fluctuations of the local field arising from the local electron spin dynamics. The nuclear relaxation rate is related to the imaginary part of the dynamic spin susceptibility $\chi''_{\perp}(\vec{q},\omega_L)$ [40]

$$\frac{1}{T_1 T} \propto \gamma_n^2 \sum_q \left|A_{\perp}(\vec{q})\right|^2 \frac{\chi''_{\perp}(\vec{q},\omega_L)}{\omega_L} \tag{4}$$

where $A_{\perp}$ is the q-dependent transverse component of the hyperfine coupling, $\vec{q}$ the wave vector, and $\omega_L$ the Larmor frequency. Thus, the $(T_1T)^{-1}$ can be considered as a measure for the local dynamic part of the local susceptibility. In a purely paramagnetic regime, there is no dependency of the susceptibility on the q-vector, and the electron spin fluctuation rate is much higher than the Larmor frequency. Therefore, the temperature dependence of the local dynamic and static susceptibility is the same, and $(T_1TK)^{-1}$ = constant, where $K$ is the line shift. The deviation of $(T_1TK)^{-1}$ from a constant value below 15 K manifests the development of electron spin correlations and a critical slowing down of the spin fluctuations (Fig. 10).

The temperature dependence of the relaxation rate in the critical regime slightly above $T_N^{NMR}$ can be described by the formula

$$T_1^{-1} \propto \left(T - T_N^{NMR}\right)^{-p} \tag{5}$$

where the fitting parameters $p$ and $T_N^{NMR}$ are found to be 1.12 ± 0.09 and $T_N^{NMR}$ = 6.7± 1 K, respectively. Using $p = \nu(z - \eta)$ [41] as well as the critical exponents $\nu$, $z$ and $\eta$ listed in Table I, we can conclude that the critical behavior of the relaxation fits well by a 3D Heisenberg or a 3D Ising model.

Table I. Values of the critical exponents $\nu$, $z$ and $\eta$ as well as the parameter $p$ expected for various spin models.

| | d | $\nu$ | $z$ | $\eta$ | $p = \nu(z - \eta)$ |
|---|---|---|---|---|---|
| MF | 3 | 0.5 [42] | 1 [42] | 0 [42] | 0.5 |
| 3D Heisenberg | 3 | 0.71 [43] | 3/2 [44] | 0.037 [43] | 1.039 |
| 3D Ising | 3 | 0.63 [45] | 2.025 [45] | 0.036 [45] | 1.253 |
| | | | 1.964 [46] | | 1.237 |
| 2D Ising | 2 | 1 [47] | 2.18 [45] | 0.25 [47] | 1.93 |
| | | | 1.75 [45] | | 1.5 |

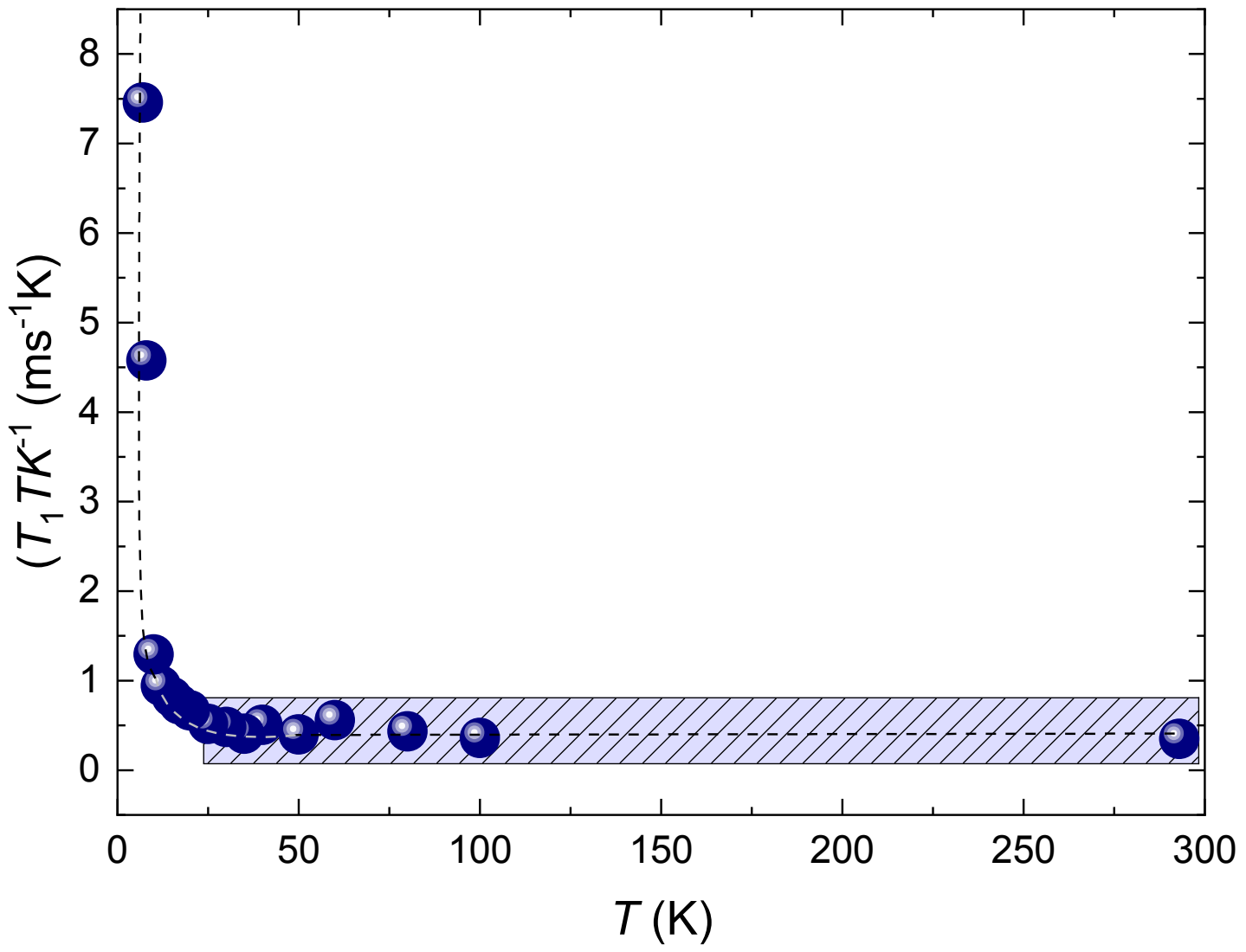

FIG. 10. The plot of $(T_1TK)^{-1}$ versus $T$, where $(T_1T)^{-1}$ represents the local dynamic susceptibility, and $K$ the local static susceptibility. The shaded area marks the region where this ratio is a constant, which corresponds to the pure paramagnetic regime.

In contrast to the $Mn^{4+}$ ESR that probes the intra-layer correlations, the NMR of the $Li^+$ ions located in between the layers should be sensitive mostly to the inter-layer interaction andconsequently to a 3D order. Thus, using a combination of two local techniques, we can clearly distinguish the temperature regions for the 2D correlations (< 150 K) and 3D correlations (15 – 6.7 K). Below $T_N^{NMR}$, the relaxation rate as well as the shape and the width of the spectrum change dramatically. The widthconsists of two contributions, namely, a rectangular shape with quadrupolar broadened wings and a relatively narrow Gaussian shape. The main part of the spectrum with rectangular shape is typical of powder samples of antiferromagnets [48]. Such a lineshape can be described by the equation

$$f(H, H_A, H_0) = \frac{1}{4H_A}\left|1 + \frac{H_0^2 - H_A^2}{H^2}\right| \tag{6}$$

for $|H_0 - H_A| \leq H \leq H_0 + H_A$. Here, $H_A$ is an absolute value of the internal magnetic field created by ion moments in the ordered phase, $H_0 = \omega_L/(\gamma_n/2\pi)$ is the external field, $\omega_L$ is a Larmor frequency, and $\gamma_n$ is the nuclear gyromagnetic ratio. We calculated the internal field on different lithium positions using the dipole point moment model and the magnetic structure proposed by neutron diffraction study (see the next section). This model is approximate, but allows us to estimate the symmetry and a possible amount of magnetically nonequivalent lithium positions in an ordered state. Our modeling results in three magnetically nonequivalent positions of lithium, so one can expect three components in the spectrum. The direction of the internal field at these positions varies by an angle of 120 degrees, but its absolute values remain the same for all three cases. For this reason, only one rectangle is observed in the powder spectrum and the internal field value obtained from the experiment corresponds to the value of $H_A$ = 1100 Oe. Comparing this value with the one resulting from the approximate model of the dipole field from point-wise spins, we can estimate the magnitude of the effective magnetic moment on the $Mn^{4+}$ ions in the ordered state to be $1.78\mu_B \pm 0.06\ \mu_B$, which is reasonable for $S = 3/2$ and $(T- T_N^{NMR})/T_N^{NMR} = 0.6$. It is interesting that the value of the internal field on $Li^+$ changes insignificantly when the external field varies from 0.78 to 7.05 T. This suggests that the external field affects mostly the in-plane components of $Mn^{4+}$ spins because, in this case, the additional magnetic fields at $Li^+$ions from two adjacent layers of $Mn^{4+}$ ions almost cancel out each other. The second relatively narrow spectral component has a low-intensity and almost vanishing line shift, which grows linearly with the external field as well as the linewidth. This contribution, probably from a small amount of nonmagnetic impurity, has long transverse and longitudinal relaxation times. Therefore, in the paramagnetic temperature region, where the time delays in the measuring sequence are short and the spectrum of the main sample is rather narrow, the relative intensity of the parasitic contribution to both the spectrum and the relaxation curves is small. At low temperatures, when the width of the main spectrum in the ordered state increases, the parasitic contribution becomes noticeable.

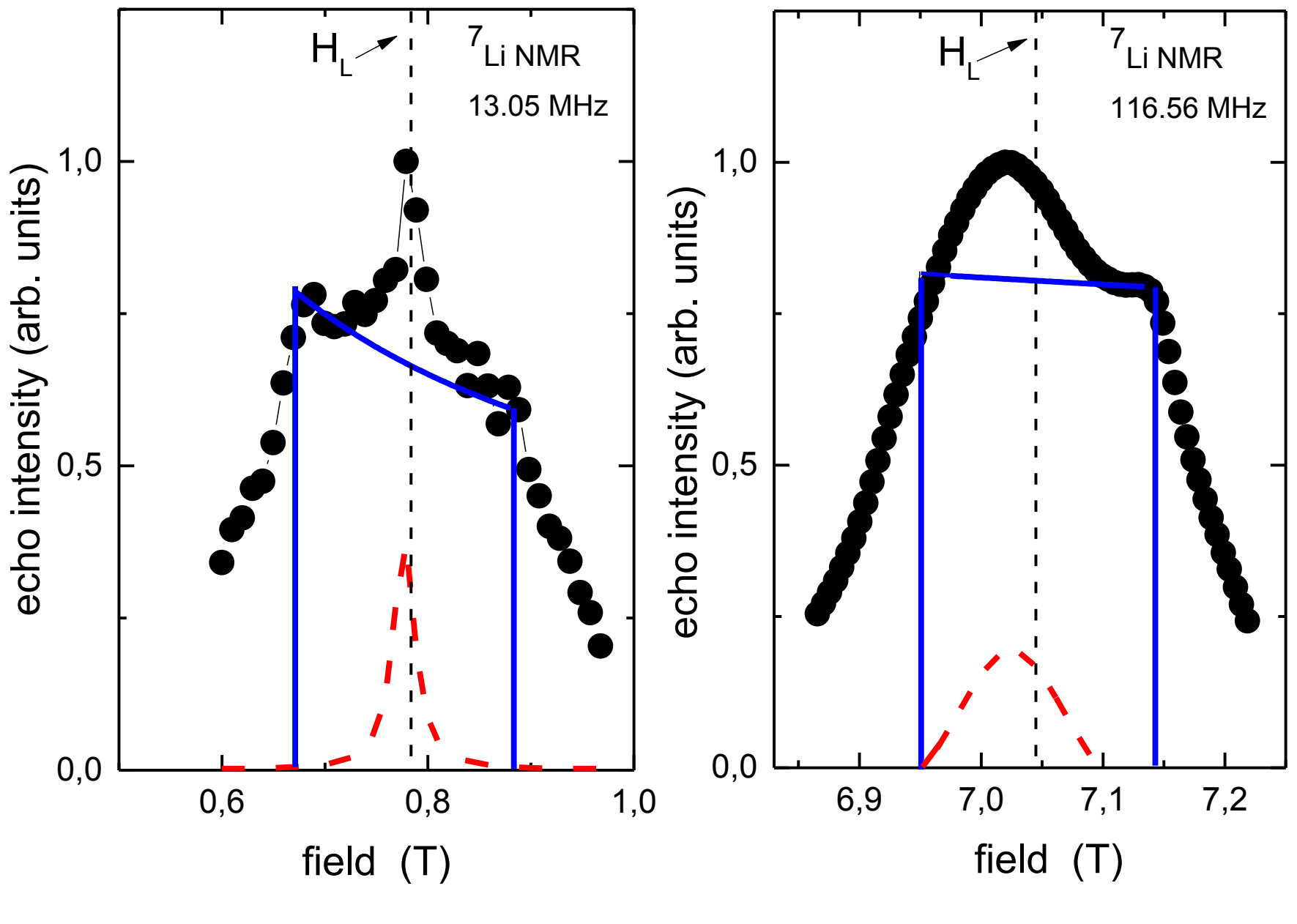

FIG. 11. The $^7$Li spectrum of $Li_2MnTeO_6$ at 4.2 K at two different frequencies (i.e., external fields). The blue lines correspond to the AF powder profiles as described by eq. (6), the reddashed lines to the impurity contribution (see the text), and the black dashed lines to the Larmor field $H_L$.

## D. Neutron powder diffraction

Fig. 12 shows NPD patterns measured on the DMC diffractometer at several low temperatures. The appearance of magnetic neutron scattering and its temperature evolution are visually demonstrated. Down to 10 K, the NPD pattern contains only nuclear reflections from the crystal lattice of the paramagnetic phase. At $T$ = 6 K, we can observe the appearance of additional reflections, including the diffraction angles smaller than the position of the first nuclear (002) reflection. The first, most intense magnetic reflections (100)-***k*** and (101)-***k*** are marked in Fig. 12. Their appearance unambiguously indicates the AFM nature of the emerging long-range spin ordering.

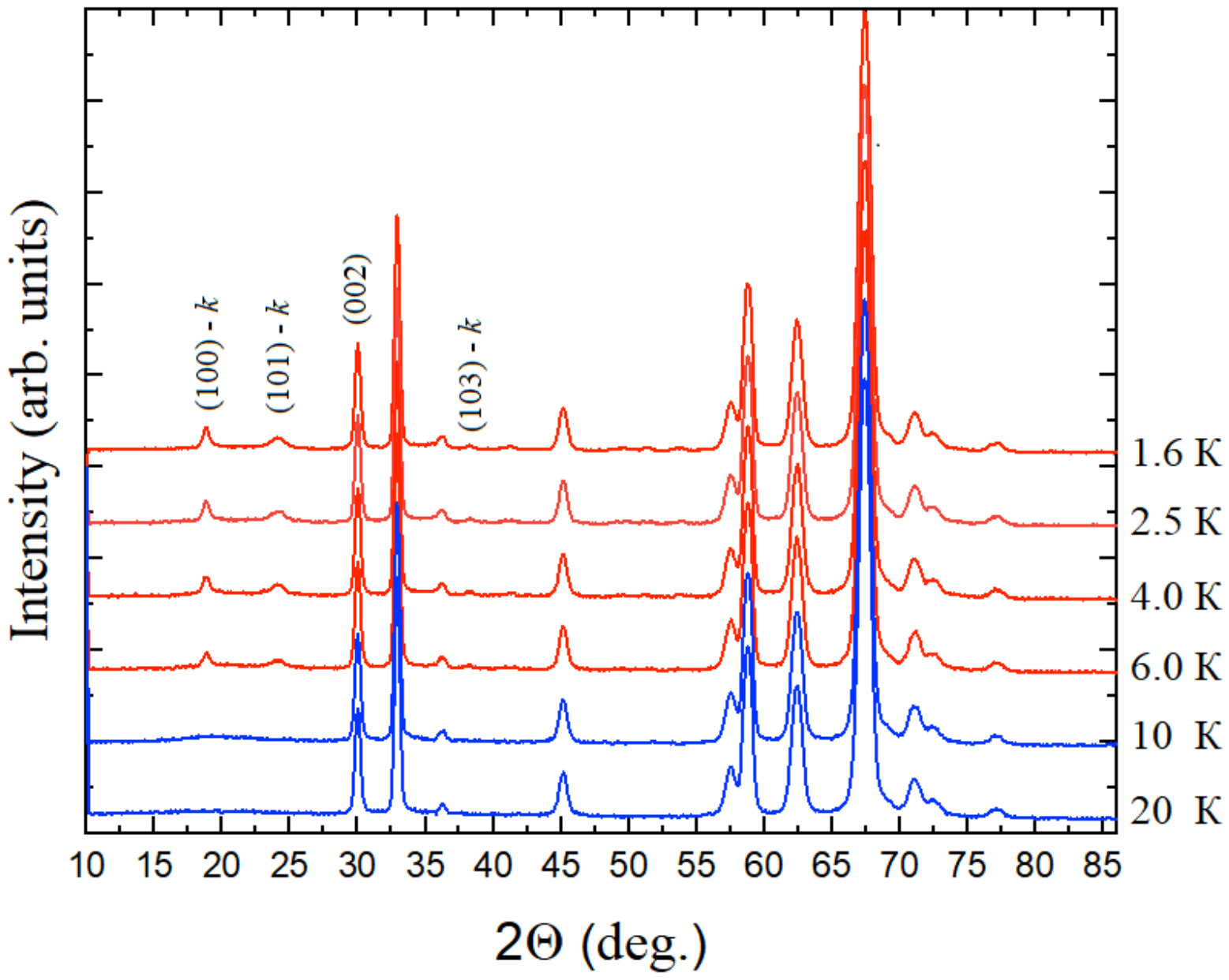

FIG. 12. Temperature dependence of the intensity of the NPD maxima measured on the DMC diffractometer at low temperatures. The indices of the most intense magnetic reflections are marked.

The value of $T_N$ estimated from the temperature dependence of the magnitude of the magnetic moment is inconsistent with that determined from the thermodynamic measurements. The data indicate a phase transition from the paramagnetic to the AFM state with decreasing temperature without any significant structural transformations. The low-temperature neutron diffraction pattern from the HRPT allowed us to construct a magnetic structure in a magnetically ordered state. Fig. 13 represents the results of the Rietveld refinement of the experimental neutron diffraction pattern measured at $T$ = 1.6 K. The appearance of the magnetic peaks is clearly seen from the inset of Fig. 13 in the 2Θ angle range between 10 and 45°. It is also worthwhile to mention that these data unambiguously demonstrate the absence of any magneticimpurity phases. The latter can manifest themselves in magnetic scattering even if these impurities are not detected by nuclear scattering due to their very low concentration.

The symmetry analysis of the seven most intensive magnetic reflections using the BasIreps program (in the FullProf_suite) shows that the observed magnetic reflections of $Li_2MnTeO_6$ are described by the propagation vector $\boldsymbol{k}$ = (1/3, 1/3, 0). This vector corresponds to a commensurate magnetic structure, and the magnetic cell is tripled along two basic crystallographic directions *a* and *b* in the layer. The magnetic structure of $Li_2MnTeO_6$ obtained from the Rietveld analysis is shown in Fig. 14. There are two antiferromagnetically ordered manganese ions per unit cell in the positions (2/3, 1/3, 1/4) and (1/3, 2/3, 3/4). In a projection view of the crystallographic unit cell on the *ab* plane, shown in the Fig. 15, two manganese ions are located with their spin moments lying in the *ab* plane and making the angle of 120° between them to forma 120° non-collinear spin arrangement in each magnetoactive $MnTeO_6$ layer with spin moment equal to 1.16$\mu_B$/Mn ($m_x$ = 0.68(2) $\mu_B$/Mn, $m_y$ = 1.38(3) $\mu_B$/Mn) (Fig. 14a). This value of the magnetic moment was obtained as a result of the full-profile processing of the neutron diffraction pattern measured at $T$ = 1.6 K. This moment is three times smaller than the effective magnetic moment calculated from the magnetic susceptibility data (Sec. IIIB) and expected for $Mn^{4+}$ ions. This reduction of the magnetic moment is very unusual. The spins in one spin triangle of one layer are pointed along the directions toward, or away from, the center of the triangle. Each triangle with spins pointed toward the center in one trigonal layer lie above and below the triangles with spins pointed away from the center (Fig. 14b).

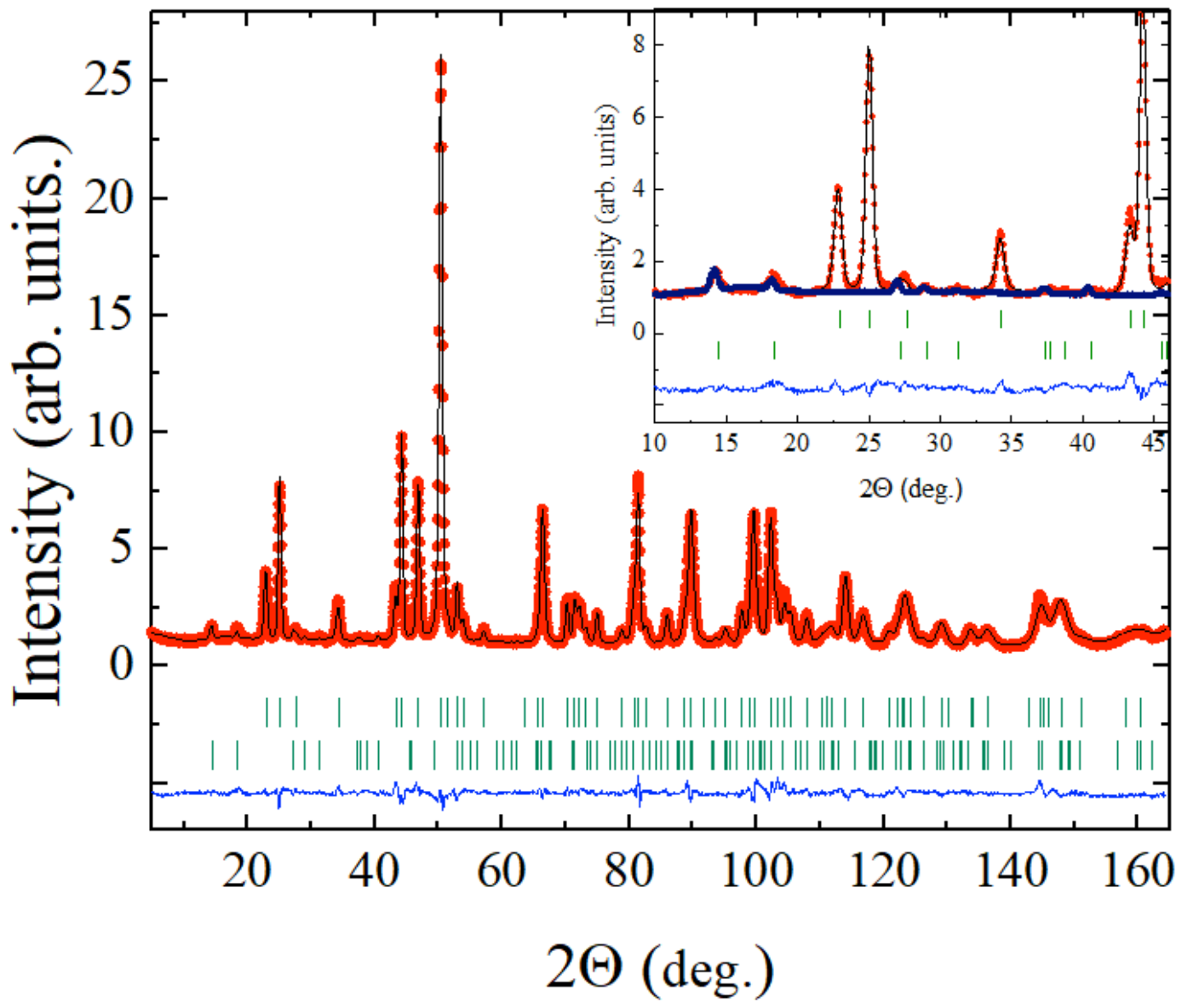

FIG. 13. Refinement of the neutron diffraction pattern of $Li_2MnTeO_6$ measured on the HRPT diffractometer at $T$ = 1.6 K. The red dots represent the experimental data, the black line shows the calculated intensity, the green ticks indicate the positions of the nuclear and magnetic Bragg reflections and the blue line shows the difference between experimental and calculated data, which is plotted at the bottom for convenience. The insert shows a small angle portion of the neutron diffraction pattern with an additional blue line showing calculated scattering from a magnetic phase.

The symmetry analysis performed with the ISODISTORT software tool [49] has shown that this magnetic structure corresponds to the maximal symmetry magnetic Shubnikov space group $P\bar{3}1c$ (165.1.1322 [50]). The group is generated by irreducible representation mK1, with special order parameter P1 (a,0), according to the international nomenclature given in Ref. [49], with the basis transformation (2,1,0),(-1,1,0), and (0,0,1). The Mn-atom is at the position 6f (x,0,1/4) with x=1/3 and the magnetic moment direction [u,0,0].

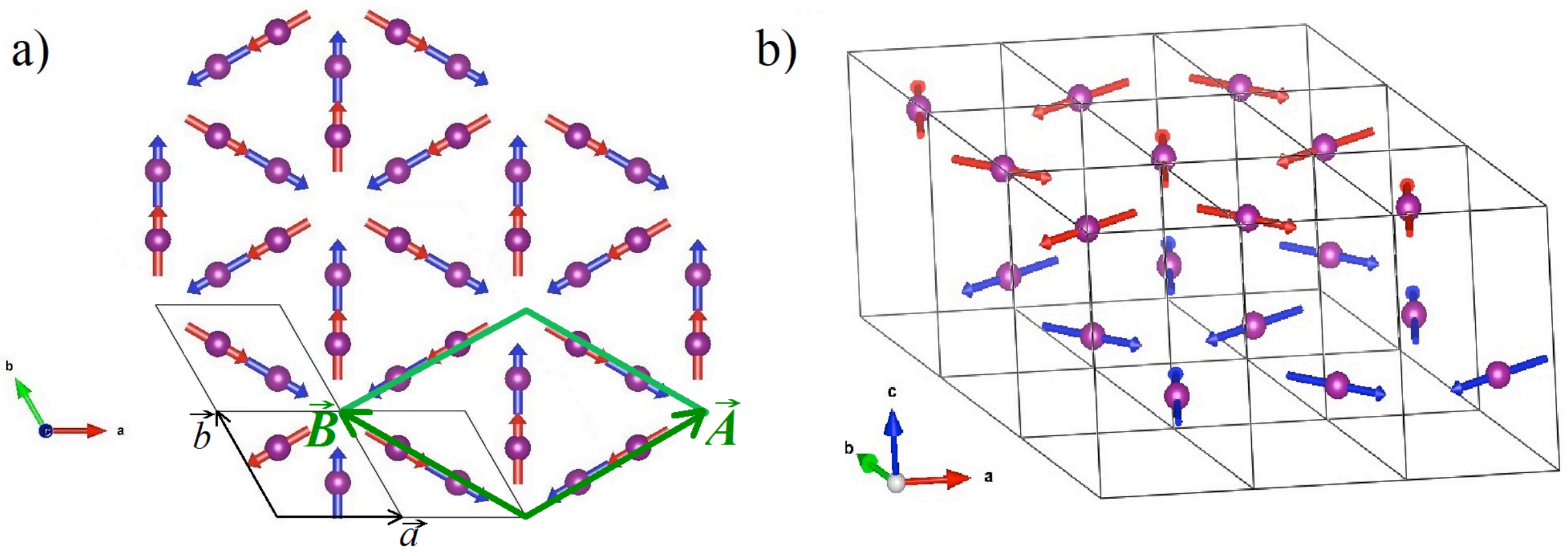

FIG. 14. Magnetic structure of $Li_2MnTeO_6$: (a) An extended projection view on the *ab*-plane of the 120° non-collinear spin arrangements of two adjacent trigonal layers (at z = 0.25 and 0.75). (b) A perspective view of the 120° non-collinear spin arrangements of two adjacent trigonal layers. The two triangles at z = 0.25 and 0.75 have a staggered arrangement. The thin black lines denote the crystallographic unit cells formed by the unit vectors **a** and **b**. The thick green lines highlight the Shubnikov magnetic unit cell formed by the unit vectors **A** and **B**. The relation between the parent crystal unit cell and the magnetic unit cell is determined by **A** = 2**a** + **b**, **B** = -**a** + **b**. ($a = b = 5.0114(2)$ Å, $c = 9.4915(9)$ Å at $T = 1.6$ K).

**E. Theoretical calculations of exchange interactions**

We evaluate the spin exchange interactions of $Li_2MnTeO_6$ by performing energy-mapping analysis based on DFT calculations [51-53]. A perspective view of the arrangement of the magnetic ions $Mn^{4+}$ in $Li_2MnTeO_6$ is presented in Figure 15, which has two triangular layers of $Mn^{4+}$ ($d^3$, $S = 3/2$) ions in a unit cell. For our analysis, we consider the nearest-neighbor spin exchange path $J_1$ within each triangular and the spin exchange path $J_2$ between adjacent triangular layers. For the evaluation of $J_1$ and $J_2$, spin-polarized DFT calculations were carried out for $Li_2MnTeO_6$ using the (2*a*, *b*, *c*) supercell containing four formula units (FUs) and three ordered spin states (FM, AF1, AF2) defined in Figure 16. Spin-polarized DFT calculations were carried out by using the Vienna ab Initio Simulation Package (VASP) [54-55], the projector augmented wave (PAW) method [56] and the PBE exchange-correlation functionals [57]. The electron correlation associated with the 3d state of Mn was taken into consideration by performing the DFT+U calculations [58] with the effective on-site repulsion $U^{eff} = U - J$. All our DFT+U calculations used the plane wave cutoff energy of 450 eV, the threshold of $10^{-6}$ eV for self-consistent-field energy convergence and the k-points of (6×6×3) set.

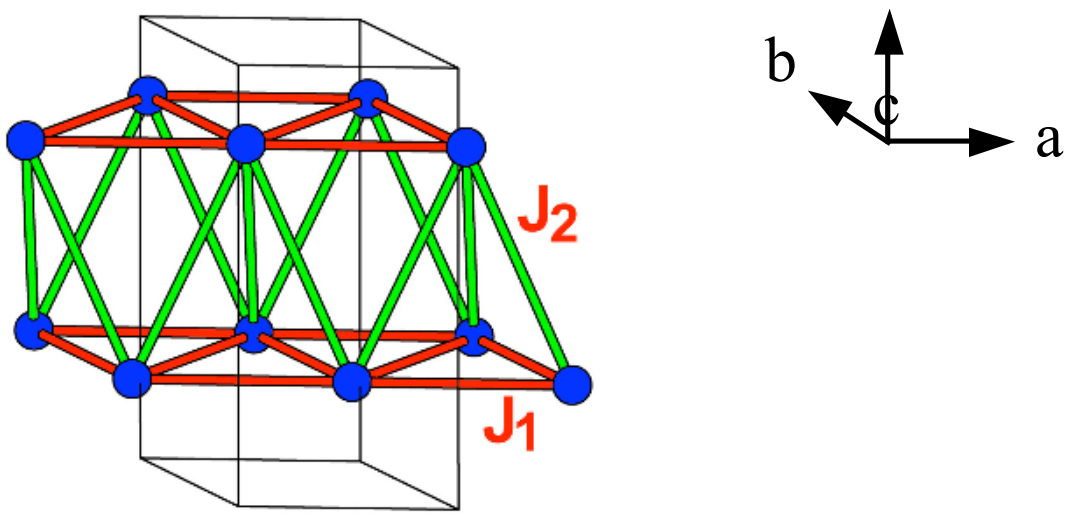

FIG. 15.Spin exchange paths $J_1$ and $J_2$ in $Li_2MnTeO_6$. The blue circles represent the $Mn^{4+}$ ions.

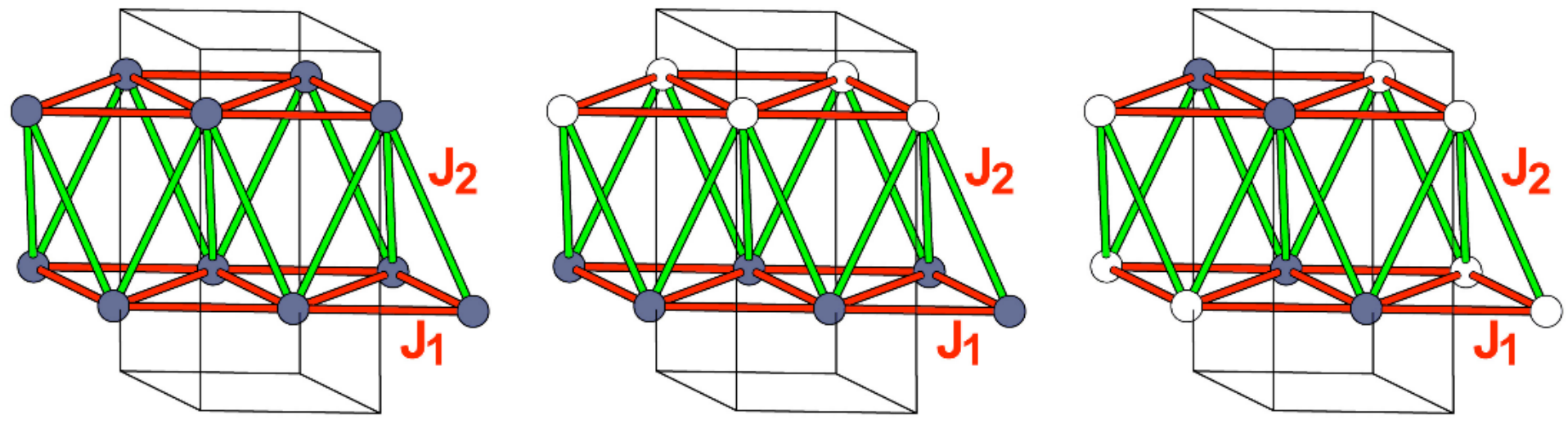

(a) FM (b) AF1 (c) AF2

FIG. 16. Spin ordered arrangements of (a) FM, (b) AF1 and (c) AF2 states, where the gray and white circles represent the up and down spin sites, respectively.

Using the spin Hamiltonian,

$$H_{spin} = -\sum_{i<j} J_{ij}\vec{S}_i \cdot \vec{S}_j, \quad (7)$$

where $\vec{S}_i$ and $\vec{S}_j$ are the spins at magnetic ion sites $i$ and $j$, respectively, with the spin exchange constant $J_{ij} = J_1, J_2$, the total spin exchange energies per four FUs of the three ordered spin states are written as

$$\begin{aligned} E_{FM} &= (-12J_1 - 12J_2)S^2 \\ E_{AF1} &= (-12J_1 + 12J_2)S^2 \\ E_{AF2} &= (4J_1 - 4J_2)S^2 \end{aligned} \quad (8)$$

where $S$=3/2, i.e., the spin of $Mn^{4+}$. The relative energies of the three ordered spin states determined by DFT+U (with $U^{eff}$ = 2, 3 and 4 eV) calculations are summarized in Table II. Thus, by mapping the relative energies onto the corresponding ones in terms of the total spin exchange energies, we obtain the values of $J_1$ and $J_2$ listed in Table II. The latter reveals that the intralayer exchange $J_1$ is AFM dominates over the interlayer exchange $J_2$ in magnitude. The latter is very weakly AFM with $U^{eff}$ = 2 eV and becomes very weakly ferromagnetic (FM) with $U^{eff}$ = 3 and 4 eV. The Curie-Weiss temperatures $\theta_{cal}$ calculated on the basis of the calculated spin exchanges $J_1$ and$J_2$ using the mean-field approximation [59] are also listed in Table II. The $\theta_{cal}$ values calculated with $U^{eff}$= 2 and 3 eV (-60 and -39 K, respectively) agree reasonably well with the experimental value of -51 ± 1 K. To a first approximation, therefore, our calculations show that the magnetic structure can be described by a triangular spin lattice with the nearest-neighbor AFM spin exchange $J_1$. The long-range magnetic order expected for a triangular spin lattice with nearest-neighbor AFM spin exchange is a 120° non-collinear spin arrangement. Therefore, our calculations are in support of our discussions presented in the previous sections.

Table II. Relative energies (in meV/FU) of the three spin ordered states, the spin exchange parameters (in K) and the Curie-Weiss temperature (in K) of $Li_2MnTeO_6$ determined from the GGA+U calculations with $U^{eff}$ = 2, 3 and 4 eV

| U(eV) | 2 | 3 | 4 |
|---|---|---|---|
| $E_{FM}$ | 6.15 | 4.04 | 2.51 |
| $E_{AF1}$ | 5.98 | 4.17 | 2.91 |
| $E_{AF2}$ | 0 | 0 | 0 |
| $J_1/k_B$ | -7.85 | -5.26 | -3.41 |
| $J_2/k_B$ | -0.15 | 0.12 | 0.34 |
| θ(K) | -60 | -39 | -23 |

## IV. CONCLUSION

In summary, the magnetic susceptibility $\chi(T)$ of the layered triangular-lattice tellurate $Li_2MnTeO_6$ reveals an unusual behavior under external magnetic field. Although $\chi(T)$ shows no obvious anomaly indicative of a long-range magnetic ordering at low magnetic fields, its character changes radically showing a maximum at about 9 K at high magnetic fields. The specific heat data exhibit a clear $\lambda$-type anomaly at $T_N \approx 8.5$ K even at zero magnetic field, unambiguously confirming that the maximum of $\chi(T)$ at high magnetic fields reflects the onset of an AFM order. This conclusion is also consistent with both $^7$Li NMR data and dielectric permittivity measurements, which indicate the presence of an AFM state below 7 K. Our DFT calculations suggest that the spins of each trigonal layer will have a 120°non-collinear spin arrangement, and that the 3D AFM ordering takes place when the interlayer spin arrangements become ordered. These suggestions are consistent with the magnetic structure determined by neutron diffraction measurements. The 3D magnetic ordering temperature $T_N \approx 8.5$ K is noticeably lower than the absolute value of the Curie-Weiss temperature $\Theta$ = -74 ± 1 K (frustration index $f = |\Theta|/T_N \approx 8$) showing the presence of strong spin frustration in each triangularspin lattice of $Li_2MnTeO_6$. Our ESR and NMR data show a 2D magnetic character in a broad temperature range above $T_N$. Support for the 2D magnetism picture in $Li_2MnTeO_6$ was also obtained from the magnetic specific heat and the critical divergence of the ESR linewidth.

## ACKNOWLEDGMENTS

The reported study was supported by Russian Foundation for Basic Research via the grant 18-02-0326 (E.A.Z., T.M.V., V.B.N. and G.V.R.) for sample preparation, magnetic, dielectric and specific heat and ESR studies; by Russian Science Foundation according to the research projects № 18-12-00375 (A.E.S. and A.I.K.) for neutron studies. NMR studies (E.V. and D.G.) were carried out with financial support from the government assignment for FRC Kazanscientific Center of RAS. The work at KHU was supported by Basic Sciences Research Program through the National Research Foundation of Korea (NRF) funded by the Ministry of Education (NRF-2017R1D1A1B03029624). The work was partially performed at the Swiss neutron spallation source SINQ at PSI. The support by the Russian Ministry of Education and Science of the Russian Federation through NUST «MISiS» grant К2-2020-008 and by the Act 211 of the Government of Russia, contracts 02.A03.21.0004, 02.A03.21.0011 is also acknowledged. A.E.S. and A.I.K. thank M. Kuchugura for her assistance in carrying out neutron measurements.

## REFERENCES

1. C. Lacroix, P. Mendels and F. Mila, *Introduction to frustrated magnetism: materials, experiments, theory* (Springer Science & Business Media, 2011), Vol. 164.
2. L. Balents, Nature (London) **464**, 199–208 (2010).
3. L. Savary and L. Balents, Rep. Prog. Phys. **80**, 016502 (2017).
4. A. Vasiliev, O. Volkova, E. Zvereva, and M. Markina, npj Quantum Materials **3**, 18 (2018).

5. N.D. Mermin and H.Wagner Phys. Rev. Lett. **17(22)**, 1133 (1966).

6. P. Chandra, P. Coleman, J.A. Mydosh, and V. Tripathi, Nature (London) **417**, 831–834 (2002).

7. F. Cricchio, F. Bultmark, O. Grånäs and L. Nordström, Phys. Rev. Lett.**103(10)**, 107202 (2009).

8. A.A.Tsirlin, A. Maisuradze, J. Sichelschmidt, W. Schnelle, P. Höhn, R. Zinke, J. Richter, and H. Rosner, Phys. Rev. B **85**, 224431 (2012).

9. T. T. M. Palstra, A. A. Menovsky, J. Van den Berg, A. J. Dirkmaat, P. H. Kes, G. J. Nieuwenhuys, and J. A. Mydosh, Phys. Rev. Lett. **55(24)**, 2727 (1985).

10. C. Broholm, H. Lin, P. T. Matthews, T. E. Mason, W. J. L. Buyers, M. F. Collins, A. A. Menovsky, J. A. Mydosh, and J. K. Kjems, Phys. Rev. B **43(16)**, 12809 (1991).

11. H. Ikeda and Y. Ohashi, Phys. Rev. Lett. **81(17)**, 3723 (1998).

12. J. P. Rodriguez, Phys. Rev. B, **82(1)**, 014505 (2010).

13. A. Zorko, P. Jeglič, A. Potočnik, D. Arčon, A. Balčytis, Z. Jagličić, X. Liu, A. L. Tchougréeff, and R. Dronskowski, Phys. Rev. Lett. **107**, 047208 (2011).

14. R. Nath, K. M. Ranjith, J. Sichelschmidt, M. Baenitz, Y. Skourski, F. Alet, I. Rousochatzakis, and A. A. Tsirlin, Phys. Rev. B **89(1)**, 014407 (2014).

15. H. D. Zhou, B. S. Conner, L. Balicas, and C. R. Wiebe, Phys. Rev. Lett. **99**, 136403 (2007).

16. M. V. Eremin, J. Deisenhofer, R. M. Eremina, J. Teyssier, D. van der Marel, and A. Loidl, Phys. Rev. B **84**, 212407 (2011).

17. I. Yamauchi, K. Nawa, M. Hiraishi, M. Miyazaki, A. Koda, K. M. Kojima, R. Kadono, H. Nakao, R. Kumai, Y. Murakami, H. Ueda, K. Yoshimura, and M. Takigawa, Phys. Rev. B **92**, 064408 (2015).

18. B. Kim, S. Khmelevskyi, P. Mohn, and C. Franchini, Phys. Rev. B **96**, 180405(R) (2017).

19. Sugiyama et.al. Phys. Rev. B **89**, 020402(R) (2014).

20. G. Jackeli and G. Khaliullin, Phys. Rev. Lett. **103**, 067205 (2009).

21. V. B. Nalbandyan, I. L. Shukaev, G. V. Raganyan, A. Svyazhin, A. N. Vasiliev, and E. A. Zvereva, Inorg. Chem., **58**, 5524−5532 (2019).

22. FULLPROF suite, http://www.ill.eu/sites/fullprof/.

23. G. A. Bain and J. F. Berry, Journal of Chemical Education, **85(4)**, 532 (2008).

24. M. E. Fisher, Proc. Roy. Soc. (London) A **254**, 66 (1960).

25. M. E. Fisher, Phil. Mag. **7**, 1731 (1962).

26. A. Tari, *The Specific Heat of Matter at Low Temperature* (Imperial College Press, London, 2003).

27. D. B Losee, J. N. McElearney, G. E. Shankle, R. L. Carlin, P. J. Cresswell, and W. T. Robinson, Phys. Rev. B **8**, 2185−2199 (1973).

28. R. L. Carlin, *Magnetochemistry* (Springer-Verlag, Berlin, 1986).

29. L. J. deJongh and A. R. Miedema, Adv. Phys. **23**, 1–260 (1974).

30. E. Vavilova, A. S. Moskvin, Y. C. Arango, A. Sotnikov, S. L. Drechsler, R. Klingeler, O. Volkova, A. Vasiliev, V. Kataev, and B. Büchner, EPL (Europhysics Letters) **88(2)**, 27001 (2009).

31. N. Terada, D. D. Khalyavin, P. Manuel, Y. Tsujimoto, K. Knight, P. G. Radaelli, H. S. Suzuki, and H. Kitazawa, Phys. Rev. Lett. **109(9)**, 097203 (2012).

32. S. Seki, Y. Onose, and Y. Tokura, Phys. Rev. Lett. **101(6)**, 067204 (2008).

33. J. Hwang, E. S. Choi, F. Ye, C. R. Dela Cruz, Y. Xin, H. D. Zhou, and P. Schlottmann, Phys. Rev. Lett. **109,** 257205 (2012).

34. J. P. Joshi and S. V. Bhat, J. of Magn. Res. **168**, 284 (2004).

35. E. Dormann and V. Jaccarino, Phys. Lett. A, **48**, 81 (1974).

36. K. Kawasaki, Prog. Theor. Phys. **39(2)**, 285–311 (1968). K. Kawasaki, Phys. Lett. A **26A**(11), 543 (1968).

37. H. Mori, K. Kawasaki, Prog. Theor. Phys. **28**(6), 971–987 (1962).

38. D. Huber, Phys. Rev. B **6**, 3180–3186 (1972).

39. A. G. Smol'nikov, V. V. Ogloblichev, A. Y. Germov, K. N. Mikhalev, A. F. Sadykov, Y. V. Piskunov, A. P. Gerashchenko, A. Y. Yakubovskii, M. A. Muflikhonova, S. N. Barilo, and S. V. Shiryaev, JETP Letters, **107(2)**, 134-138 (2018).

40. T. Moriya, J. Phys. Soc. Jpn. **18**, 516 (1963).

41. F. Borsa, M. Corti, T. Goto, A. Rigamonti, D. C. Johnston, F. C. Chou, Phys. Rev. B **45**, 5756– 5759 (1992).

42. H. E. Stanley, *Introduction to phase transitions and critical phenomena* (Oxford University Press, 1987).

43. M. Campostrini, M. Hasenbusch, A. Pelissetto, P. Rossi, and E. Vicari, Phys. Rev. B**65(14)**, 144520 (2002).

44. P. C. Hohenberg and B. I. Halperin, Rev. Mod. Phys. **49**, 435 (1977).

45. A. Pelissetto and E. Vicari, Phys. Rep. **368**, 549 (2002).

46. L. Van Hove, Phys. Rev. **95**, 249 (1954).

47. L. Onsager, Crystal statistics. I. Phys. Rev. **65**, 117 (1944).

48. Y. Yamada and A. Sakata, J. Phys. Soc. Jpn. **55**, 1751 (1986).

49. B.J. Campbell, H.T. Stokes, D.E. Tanner, and D.M. Hatch, J. Appl. Crystallogr.**39**, 607–614 (2006) (also available via the Internet at: http://iso.byu.edu/iso/isotropy.php, ISOTROPY Software Suite).

50. D.B.Litvin,*1-, 2- and 3-Dimensional Magnetic Subperiodic Groups and Magnetic Space Groups* (Chester: International Union of Crystallography, 2013).

51. M.-H. Whangbo, H.-J. Koo, and D. Dai, J. Solid State Chem. **176**, 417 (2003).

52. H.J. Xiang, C. Lee, H.-J. Koo, X.G. Gong, and M.-H. Whangbo, Dalton Trans. **42**, 823 (2013).

53. M.-H. Whangbo, H.J. Xiang, *Magnetic Properties from the Perspectives of Electronic Hamiltonian: Spin Exchange Parameters, Spin Orientation and Spin-Half Misconception*, in Handbook in Solid State Chemistry, Volume 5: Theoretical Descriptions, Dronskowski, R., Kikawa, S., Stein, A., eds., Wiley, 2017, 285-343.

54. G. Kresse and J. Hafner, Phys. Rev. B **47**, 558 (1993).

55. G. Kresse and J. Furthmüller, Comput. Mater. Sci. **6**, 15 (1996).

56. G. Kresse and J. Furthmüller, Phys. Rev. B **54**, 11169 (1996).

57. J. P. Perdew, K. Burke, and M. Ernzerhof, Phys. Rev. Lett. **77**, 3865 (1996).

58. S. L. Dudarev, G. A. Botton, S. Y. Savrasov, C. J. Humphreys, and A. P. Sutton, Phys. Rev. B **57**, 1505 (1998).

59. J. S. Smart, *Effective Field Theory of Magnetism*, Saunders, Philadelphia, 1966.